\documentclass[10pt,amsmath,amssymb,twocolumn,superscriptaddress,nofootinbib,aps,prd,twocolumn,floatfix]{revtex4-2}

\usepackage{graphicx}  % needed for figures
\graphicspath{{./CME_plots/}}
\usepackage{bm}        % for math
\usepackage{enumitem}
\usepackage{etoolbox}

\usepackage[colorlinks=true]{hyperref}

\providecommand{\texorpdfstring}[2]{#1}

\newcommand{\comment}[1]{}

\newcommand{\lr}[1]{ \left( #1 \right) }
\newcommand{\lrs}[1]{ \left[ #1 \right] }

\newcommand{\vev}[1]{ \langle \, #1 \, \rangle }

\newcommand{\abs}[1]{ \left| #1 \right| }

\newtoggle{twocolumn}
\toggletrue{twocolumn}

\begin{document}
	\sloppy
	
	\title{Chiral Magnetic Effect and Negative Magnetoresistance\\ across the phase diagram of finite-density \texorpdfstring{$\boldmath SU(2)$}{SU(2)} gauge theory}
	
	\author{P.~V.~Buividovich}
	\email{pavel.buividovich@liverpool.ac.uk}
	\affiliation{Department of Mathematical Sciences, University of Liverpool, Liverpool, L69 7ZL, UK}
	%\affiliation{\UL}
	
	\author{L.~{von Smekal}}
	\email{lorenz.smekal@physik.uni-giessen.de}
	\affiliation{Institut f\"ur Theoretische Physik, Justus-Liebig-Universit\"at, 35392 Giessen, Germany}
	\affiliation{Helmholtz Research Academy Hesse for FAIR (HFHF), Campus Giessen, 35392 Giessen, Germany}

    \author{D.~Smith}
	\email{d.smith@gsi.de}
	\affiliation{GSI - Helmholtzzentrum für Schwerionenforschung, 64291 Darmstadt, Germany}
		
	\date{August 6th, 2026}
	
\begin{abstract}
	We study the signatures of the Chiral Magnetic Effect (CME) in $SU\lr{2}$ gauge theory with $N_f = 2$ flavours of dynamical fermions at finite temperature $T$, quark chemical potential $\mu$ and a weak external magnetic field $e B$. We consider both the correlator of the axial density and the vector current, which gives direct access to the out-of-equilibrium CME, and the correlator of two vector currents, which probes the CME indirectly via the enhancement of the longitudinal electric conductivity (Negative Magnetoresistance, NMR). We find that the CME response extracted from the vector-axial correlator exhibits a rather weak dependence on temperature and density in the quark-gluon plasma regime, and is very close to the universal value for free massless quarks. The CME appears to be mildly suppressed at low temperatures in the hadronic phase. In contrast, the NMR behaves in a qualitatively different way across the phase diagram, and is strongly suppressed at either large densities or temperatures. The magnitude of the NMR response appears to be considerably smaller than the prediction based on the lowest Landau level calculation for free quarks. Our findings suggest that for relatively small magnetic field strengths $e B \lesssim m_{\pi}^2$ the relation between the CME and NMR might not be as direct as expected. We also do not find statistically significant indications for an enhancement of the CME or NMR strength in the vicinity of the crossover or second-order phase transition lines in the $\lr{\mu, T}$ phase diagram.	
\end{abstract}
	
\maketitle
	
\section{Introduction}
\label{sec:intro}

The experimental search for statistically significant signatures of the Chiral Magnetic Effect (CME) in heavy-ion collisions is currently an active research area. After the initial non-observation of the CME in a blind analysis of the data obtained in the isobar run at RHIC \cite{STAR:2109.00131,STAR:2209.03467}, the status of the CME in heavy-ion collisions remained controversial. Post-blind analyses \cite{Kharzeev:2022hqz,Lacey:2022plw} suggest that experimental data might still be compatible with a non-vanishing CME signal. A lot of work was done on optimizing the experimental observables, event-shape selection techniques, and background suppression \cite{Li:2024gdz,Guo:2025wry,Feng:2502.09742,Li:2511.07358}, culminating in the recent report of a statistically significant CME signal in a limited range of beam energies in RHIC Beam Energy Scan II \cite{STAR:2506.00275}.

The fact that the CME signal seems to be detectable only for specific beam energies (and probably isotope/isobar numbers) suggests that the CME strength might vary significantly across the phase diagram of strongly interacting matter \cite{STAR:2506.00275}. For example, model calculations in \cite{Ikeda:2012.02926} and functional renormalization group calculations in \cite{Fejos:2506.14010} suggest that chirality fluctuations driving the CME might be enhanced in the vicinity of a critical point. This observation clearly calls for first-principle estimates of the CME strength across the phase diagram of finite-density, hot QCD. In particular, the vicinity of the critical endpoint at finite temperature and density is the target region for ongoing heavy-ion collision experiments. The planned experiments at NICA and FAIR facilities will achieve even larger baryon densities at lower temperatures.

However, not much is known about the CME response across the phase diagram of finite-density QCD. Most theoretical studies focused on an equilibrium setup where dynamical chirality fluctuations are modeled in terms of a finite ``chiral chemical potential'' $\mu_5$ \cite{Gynther:1005.2587,Rebhan:0909.4782}. Because of the fermion sign problem at finite baryon density, numerical studies of the CME within the lattice QCD framework were likewise only performed at zero baryon chemical potential \cite{Buividovich:09:7,Yamamoto:1111.4681,Yamamoto:1105.0385,Brandt:2405.09484}. 

\begin{figure}[h!tpb]
	\includegraphics[width=0.49\textwidth]{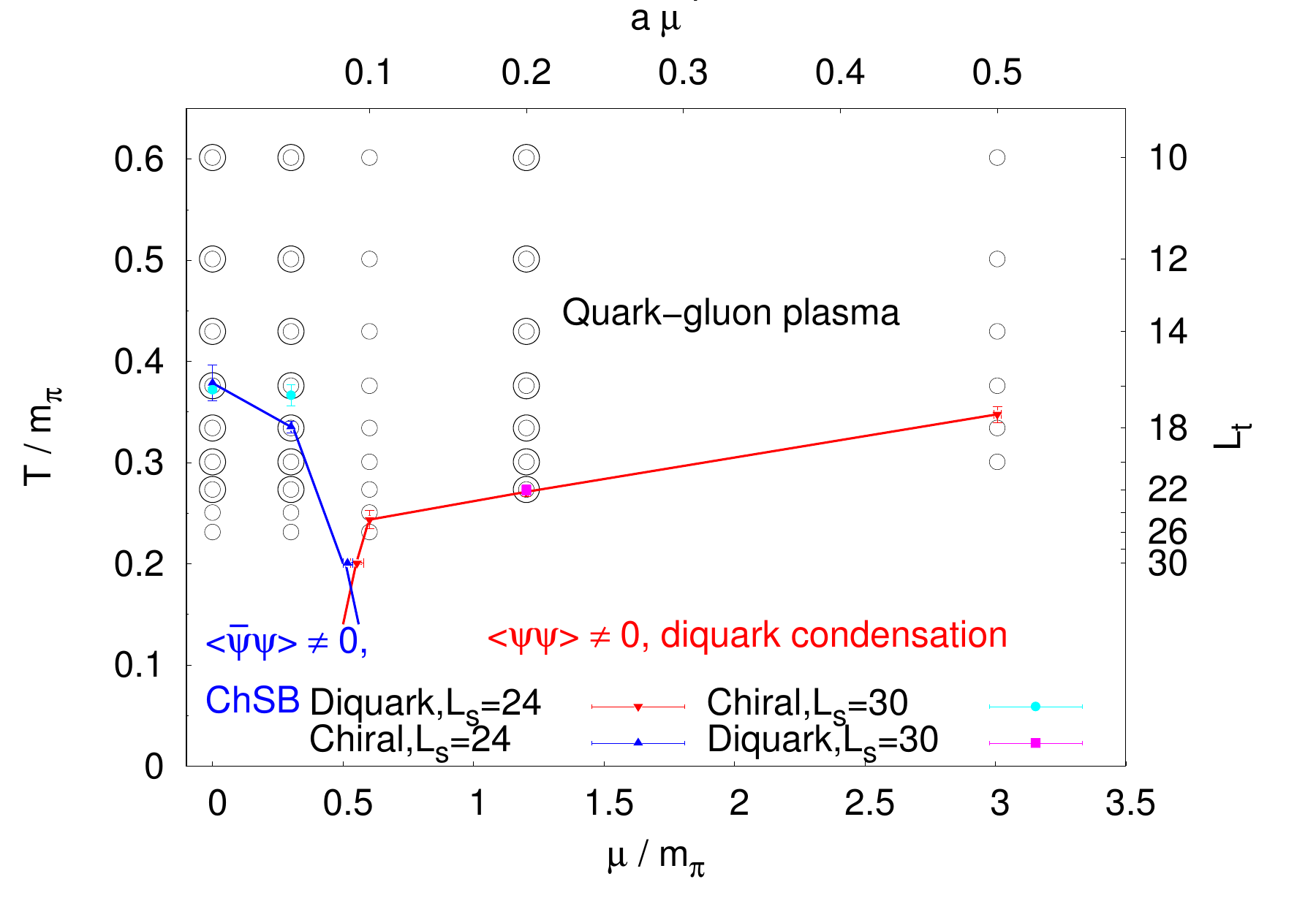}
	\caption{Phase diagram of $SU(2)$ lattice gauge theory with $N_f = 2$ light quark flavours in the $\lr{T, \mu}$ plane, obtained using the same set of gauge field configurations and same parameters as in the present work. Figure adapted from our work \cite{Buividovich:20:1}. Blue line is a chiral crossover and red line is a 2nd order phase transition \cite{Braguta:1605.04090,Smith:1910.04495,Lawlor:2210.07731}.}
    \label{fig:phase_diagram}
\end{figure}

While a direct numerical study of the CME in finite-density lattice QCD might still be impossible, it still makes sense to study the CME in finite-density gauge theories which resemble real QCD in one way or another. $SU(2)$ gauge theory is probably one of the simplest {non-Abelian} gauge theories which can be simulated at finite fermion density and still shares many important features with real QCD \cite{Kogut:hep-lat/0105026,Kogut:hep-ph/0001171,Suenaga:2606.13974}. The phase diagram of finite-density $SU(2)$ lattice gauge theory is well studied by now and we refer the reader to numerous existing literature \cite{Hands:hep-lat/0604004,Hands:1001.1682,Smekal:1112.5401,Hands:1210.4496,Strodthoff:1306.2897,
Hands:1502.01219,Braguta:1605.04090,Braguta:1711.01869,Holicki:1701.04664,Smith:1910.04495,Etou:1910.07872,Huber:1909.12796,Hands:1912.10975,Lawlor:2210.07731} as well as Fig.~\ref{fig:phase_diagram} (adapted from our work \cite{Buividovich:20:1}). At sufficiently high temperatures, the phase diagram features the quark-gluon plasma regime. At low temperatures and small values of the quark chemical potential $\mu < m_{\pi}/2$ there is a hadronic regime with spontaneously broken chiral symmetry, qualitatively similar to the hadronic regime of QCD. These two regimes are separated by a chiral crossover, similar to the one in real QCD. In real QCD the hadron gas at low $T$ extends all the way to the nuclear liquid-gas transition, at a value of the quark chemical potential which is given by one third of nucleon mass minus binding energy per nucleon of symmetric nuclear matter, i.e.~up to $\mu \approx 308$~MeV. In contrast, in $SU(2)$ gauge theory baryons are diquarks which have the same mass as the pions and hence condense much earlier, at $\mu = m_{\pi}/2$ in the low-temperature limit. Diquark condensation is a second-order phase transition, shown as a red line in Fig.~\ref{fig:phase_diagram}. At higher temperatures it shifts towards larger values of $\mu$, goes through the chiral crossover region, and continues to higher densities and temperatures while still being a second-order transition \cite{Braguta:1605.04090,Smith:1910.04495,Lawlor:2210.07731}. 

Because of diquark condensation, $SU(2)$ gauge theory can only be reliably compared with real QCD at sufficiently small $\mu$. At very large densities and low temperatures, $SU(2)$ gauge theory may also exhibit similarities with the conjectured quarkyonic phase in QCD \cite{Pisarski:0706.2191,Braguta:1605.04090}. We also note that the physics of $SU(2)$ gauge theory is very similar to the physics of QCD with isospin chemical potential \cite{Endrodi:1407.1216,Smekal:2502.04025}, where one can derive exact inequalities between physical quantities at finite isospin and at finite baryon chemical potentials \cite{Cohen:hep-ph/0304024}.

Despite the diquark condensation transition being in a different universality class from the QCD critical endpoint ($O(2)$ vs.\ Ising universality), it still features a diverging correlation length and enhanced order parameter fluctuations \cite{Braguta:1605.04090,Smith:1910.04495,Lawlor:2210.07731,Buividovich:20:1}. Thus one can also hope that finite-density $SU(2)$ gauge theory will allow to learn some general lessons about the behavior of CME in the vicinity of a chiral crossover or a possible second-order phase transition with a QCD critical point at finite density. 

In this paper, we report on a systematic study of the CME and its main proxy effect, Negative Magnetoresistance \cite{Fukushima:0912.2961,Buividovich:10:1,Braguta:2406.18504,Braguta:1910.08516,Kharzeev:1412.6543,Sun:1603.02624,Fukushima:2106.07968,Braguta:1707.09810}, in $SU(2)$ lattice gauge theory with $N_f = 2$ flavours of light quarks at finite temperature and density. First, we use the recently introduced $CP$-odd Euclidean-time correlator $\vev{Q_A\lr{0} j\lr{\tau}}$ of axial charge and electric current in an external magnetic field \cite{Buividovich:24:2,Brandt:2502.01155} to extract the out-of-equilibrium CME response from Euclidean-time lattice simulations. 

However, as the local axial charge imbalance that drives the CME is a $\mathcal{CP}$-odd quantity and not directly observable in experiment, most experimental searches focus on $\mathcal{CP}$-even observables that are sensitive to preferential emission of charged hadrons in the direction perpendicular to the reaction plane \cite{Kharzeev:1908.07605}. Most such observables can be related to correlators of electric currents in the hot QCD medium \cite{Voloshin:hep-ph/0406311,Fukushima:0912.2961,Magdy:1710.01717,Yin:1504.06906}. In terms of these observables, the CME manifests itself as Negative Magnetoresistance, the enhancement of electric current fluctuations and electric conductivity in the direction of the magnetic field. In addition to the CME itself, we therefore also consider the dependence of the electric conductivity on magnetic field and fermion density.

A limitation of our study is that the magnetic field is only introduced for valence quarks (that is, in the current-current correlators) and does not affect virtual fermion loops (that is, the fermion determinant in the path integral weight). The reason is that an external magnetic field breaks the time-reversal symmetry that guarantees the absence of the fermionic sign problem for the (pseudo-real) $SU(2)$ gauge group. As a result, simulations with both, finite density and external magnetic field will suffer from  a complex-valued fermionic determinant even for $SU(2)$ gauge theory. While for sufficiently small magnetic fields this problem could potentially be solved by re-weighting or Taylor expansion in powers of magnetic field strength \cite{Endrodi:1407.1216}, this will significantly increase the computational cost and simulation complexity. We therefore pursue a simpler ``magnetic-quenched'' approach in this work, adding small uniform magnetic fields only for the quark propagators that are used to calculate the correlators of vector and axial currents. With only the smallest possible non-zero values $\Phi = 1, 2$ of the magnetic flux $\Phi$, one can think of our results as estimates of coefficients in a Taylor expansion of conductivity and CME response in powers of an external magnetic field $\vec{B}$ around $\vec{B} = 0$ \cite{Endrodi:1407.1216}, rather than simulations which probe how strong magnetic fields change the phase diagram of the theory. Expanding the fermionic determinant in the external magnetic field, one can see that any corrections to current-current correlators that come from the determinant correspond to disconnected fermionic diagrams with electric current insertions whose contributions to the electric conductivity are usually suppressed \cite{Buividovich:20:1}. Therefore, for not too large magnetic fields, which cannot cause the theory to transition to another phase,  we expect the effects of magnetic quenching to be reasonably small.

The main conclusion of our study is that finite chemical potential, either in the vicinity of a chiral crossover and/or second-order phase transition, or away from them, neither causes a significant suppression nor a significant enhancement of the ``pure'' CME signal as extracted from the correlator of axial charge and vector current in the direction of the magnetic field. Negative Magnetoresistance, in contrast, is found to be significantly suppressed at finite fermion density. The relative sensitivity of the longitudinal electric conductivity $\sigma_{\parallel}$ to the magnetic field, which we measure in terms of the ratio $\frac{1}{\sigma_{\parallel}} \frac{\partial^2 \sigma_{\parallel}}{\partial \lr{e B}^2}$, does not significantly exceed its zero density limit for chemical potentials up to approximately $\mu \lesssim 3 \, m_{\pi}$. We do not find statistically significant evidence for an enhancement of either the CME or NMR, neither in the chiral crossover region nor across the phase boundary of the second-order diquark condensation transition in the $\mu-T$ plane.

We also point out that even at zero density, the correlators of electric currents $\vev{j_i\lr{0} j_k\lr{\tau}}$ are much less sensitive to external magnetic fields than the axial-vector correlator $\vev{Q_A\lr{0} j_k\lr{\tau}}$ (we assume volume-averaging or, equivalently, zero spatial momentum). For free fermions in external magnetic fields, both correlators receive the same contribution from the lowest Landau level. Higher Landau levels do not contribute to the axial-vector correlator. The results reported in \cite{Buividovich:24:2} and in this paper suggest that the lowest Landau level contribution is also not changed significantly due to interactions (see \cite{zhang2019hall,zhang2020influence,zubkov2023effect} for closely related proofs of interaction independence). On the other hand, the vector-vector correlator receives interaction-dependent contributions from all Landau levels, which makes it a much more contaminated observable for detecting the CME. 

The structure of the paper is the following: in Section~\ref{sec:lattice_setup} we discuss our lattice setup and the technical details of our simulations. In Section~\ref{sec:CME} we directly study the non-equilibrium CME response at finite temperature and density in terms of the correlator of axial charge and vector currents. In Section~\ref{sec:NMR} we consider the electric conductivity at finite density and in external magnetic fields, and discuss the Negative Magnetoresistance phenomenon. We conclude in Section~\ref{sec:conclusions}. 

\section{Lattice setup and simulation details}
\label{sec:lattice_setup}

We use the same lattice setup and set of configurations as in \cite{Buividovich:20:1,Buividovich:20:2,Buividovich:21:1}. Gauge field configurations are generated using $N_f=2$ mass-degenerate rooted staggered fermions and a tree-level improved Symanzik gauge action implemented within a standard HMC algorithm. Spatial lattice sizes for our gauge field ensembles are $L_s = 24$ and $L_s = 30$, temporal sizes are $L_t = 10, 12, 14, 16, 18, 20, 22, 24, 26$. We use a single value of the bare gauge coupling $1/g^2 = 1.7$ which offers a reasonable balance between strong-coupling lattice artifacts \cite{Scheffler:1311.4324} and large lattice volumes in physical units. Correspondingly, we work with a single value of lattice spacing $a$. We did not fix $a$ in the standard way using the string tension and mainly work with dimensionless, scale-independent ratios like $T/m_{\pi}$ and $\mu/m_{\pi}$. The quark chemical potential takes values $a \mu = 0.0, \, 0.05, \, 0.1, \, 0.2, \, 0.5$ and $a \mu = 0.0, \, 0.05, \, 0.20$ in lattice units for $L_s = 24$ and for $L_s = 30$, respectively. We also use a small diquark source term with coefficient $a \lambda = 5 \cdot 10^{-4}$ as a seed for a diquark condensation transition, which would otherwise be difficult to control in a finite volume \cite{Kogut:hep-lat/0104010}. 

The zero-temperature, zero-density pion mass (in lattice units) is $a \, m_{\pi} = 0.158 \pm 0.002$ for our ensembles, with a reasonably small but not yet physical ratio between the pion and the $\rho$-meson masses $m_{\pi}/m_{\rho} \approx 0.4$. 

The phase diagram of $SU(2)$ lattice gauge theory, with parameters as outlined above, is summarized in Fig.~\ref{fig:phase_diagram}. Double concentric circles in this phase diagram correspond to ensembles with both $L_s = 24$ and $L_s = 30$ spatial lattice sizes, and single circles to ensembles with $L_s = 24$ only. The phase boundaries were estimated in \cite{Smith:1910.04495,Buividovich:20:1} using chiral and diquark susceptibilities. 

The axial-vector and vector-vector correlators are measured using the Wilson-Dirac operator with HYP smearing. The bare mass in the Wilson-Dirac operator is tuned to match the pion mass $a \, m_{\pi} = 0.158$ measured with Wilson-Dirac and with staggered valence quarks \cite{Renner:hep-lat/0409130,Edwards:hep-lat/0510062,Berkowitz:1704.01114}. We use conserved vector currents and a point-split axial charge \cite{Rakow:1511.05304,Rakow:1612.04992} as in \cite{Buividovich:20:1,Buividovich:20:2,Buividovich:21:1,Buividovich:24:2}. The ease of constructing such currents and the absence of any artificial differences between even and odd time slices are our main reasons for using Wilson-Dirac valence quarks on staggered sea quarks in a mixed-action setup. The multiplicative renormalization factor $Z_A$ for the axial current was estimated as $Z_A = 1.1 \pm 0.05$ in \cite{Buividovich:24:2}, and is taken to be $Z_A = 1.1$ in what follows.  

\section{Chiral Magnetic Effect at finite fermion density}
\label{sec:CME}

While most theoretical and numerical studies of the CME use a ``chiral chemical potential'' $\mu_5$ to induce chirality imbalance, it was recently argued in \cite{Buividovich:24:2,Brandt:2502.01155} that the proper way to characterize the non-equilibrium CME response is to consider the correlators of the spatial vector current $j_k = \sum_f q_f \, \bar{\psi}_f \, \gamma_k \, \psi_f$ in the direction of the external magnetic field $\vec{B}$ with the axial charge $Q_A = \int d^3 \vec{x} \sum_f \bar{\psi}_f \, \gamma_5 \, \gamma_0 \, \psi_f$. In real time, this correlator can be directly related to the real-time axial anomaly equation. For free fermions in Euclidean (imaginary) time, this correlator only receives contributions from the lowest Landau level and takes the following form:
\begin{eqnarray}
\label{eq:CME_Euclidean_correlator}
	\frac{1}{V} \int d^3 \vec{x} \,\vev{Q_A\lr{0} j_k\lr{\tau, \vec{x}}} 
	= \nonumber \\ = 
	\frac{N_c \, C_{em} \, e B_k}{2 \pi^2} \lr{T - \delta\lr{\tau}} ,
\end{eqnarray}
where $V$ is the total spatial volume of the system, and $\tau \in \lrs{0, T^{-1}}$ is the Euclidean time. $N_c$ is the number of quark colours, $C_{em} = \sum_f q_f^2$ the sum of squared quark charges over all light quark flavours (which we assume mass-degenerate), and $B_k$ is the external magnetic field.

\begin{figure*}[h!tpb]
	\includegraphics[width=0.49\textwidth]{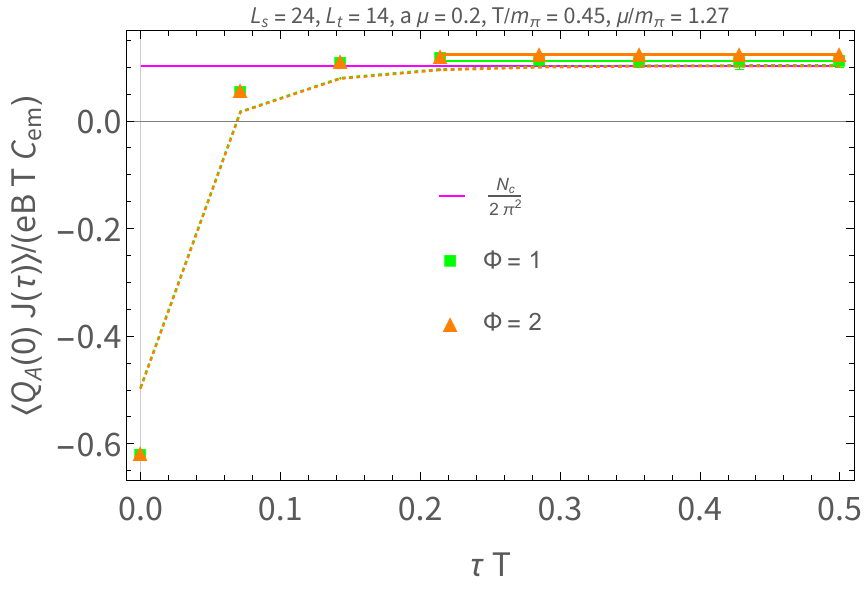}
	\includegraphics[width=0.49\textwidth]{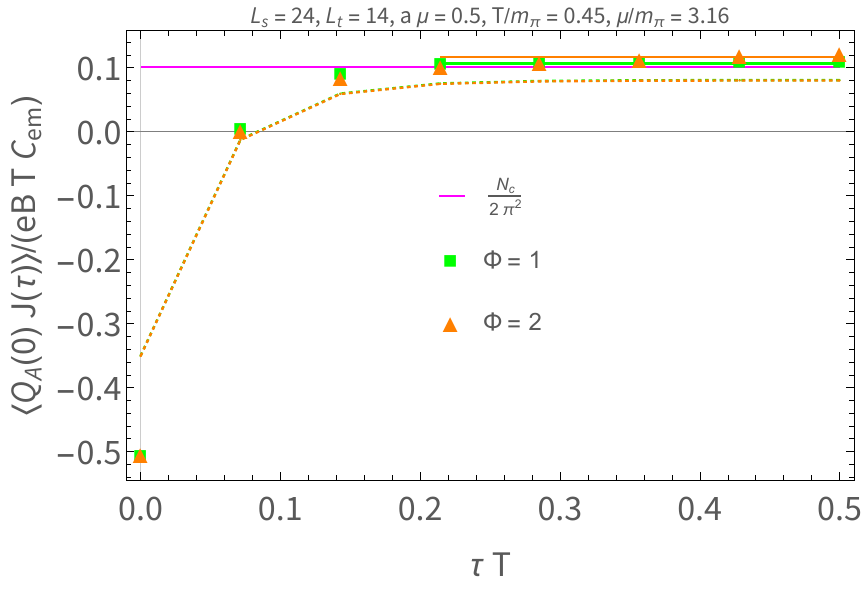}\\
	\includegraphics[width=0.49\textwidth]{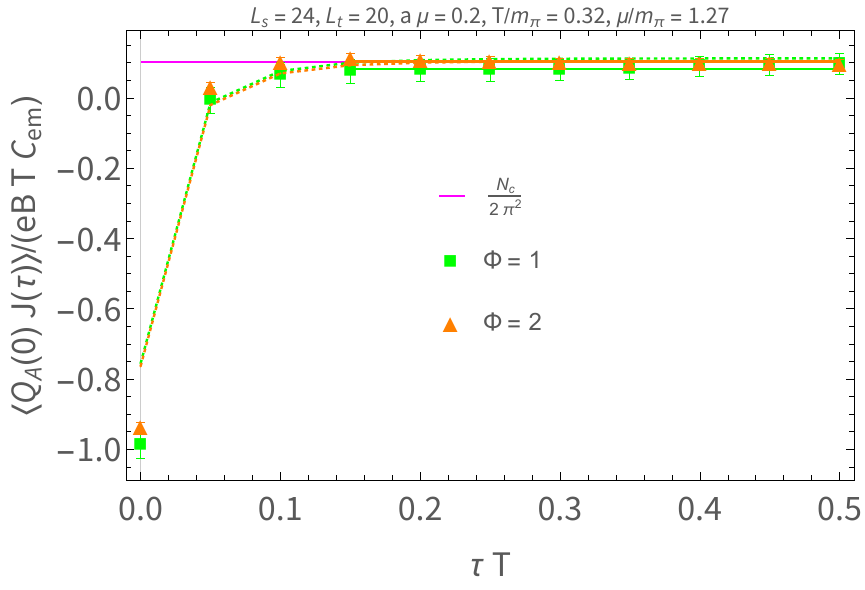}
	\includegraphics[width=0.49\textwidth]{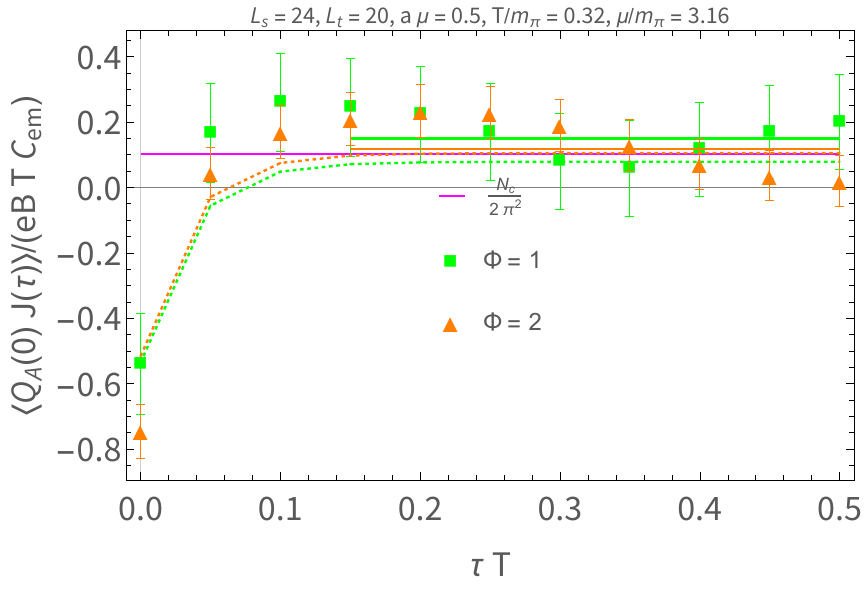}
	\caption{Axial-vector correlators $V^{-1} \int d^3 \vec{x} \vev{Q_A\lr{0} j_k\lr{\tau, \vec{x}}}$  as functions of Euclidean time $\tau$ at different values of temperature, fermion chemical potential, and magnetic field strength $e B = \frac{2 \pi \Phi}{L_s^2 a^2}$, where $\Phi$ is the total magnetic flux. Data points with error bars and dashed lines correspond to the full gauge theory case and the free lattice quarks, respectively. Horizontal shaded bars correspond to the average of the axial-vector correlator over Euclidean time $\tau$ in the range $\lrs{4 \, a, \beta - 4 \, a}$.}
	\label{fig:QAJV}
\end{figure*}

Lattice results for the axial-vector correlator are shown on Fig.~\ref{fig:QAJV}. As discussed in \cite{Buividovich:24:2} for the case of finite temperature and zero density, even for the full gauge theory the axial-vector correlator is quite close to the free-fermion, continuum result (\ref{eq:CME_Euclidean_correlator}). Namely, there is a broad and flat plateau with the height that is very close to the free quark value $\frac{N_c \, C_{em} \, \lr{e B} \, T}{2 \pi^2}$ in (\ref{eq:CME_Euclidean_correlator}) at intermediate values of Euclidean time separation $\tau$ (magenta lines in Fig.~\ref{fig:QAJV}), along with negative-valued contact terms that are localized within two-three lattice spacings around the endpoints $\tau = 0$ (identified with $\tau = T^{-1}$) of the Euclidean time interval.  

\begin{figure*}[h!tpb]
	\includegraphics[width=0.49\textwidth]{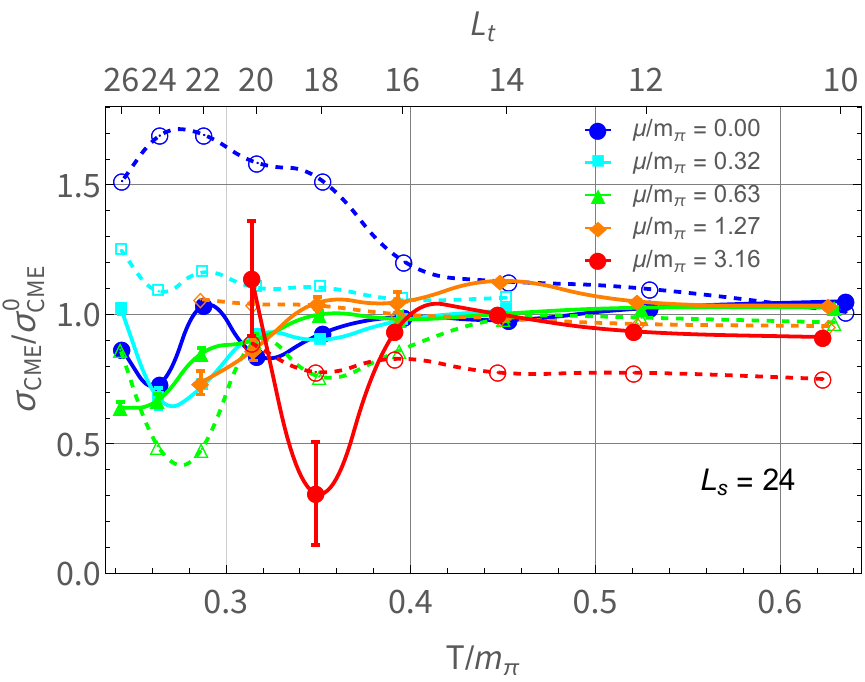}\includegraphics[width=0.49\textwidth]{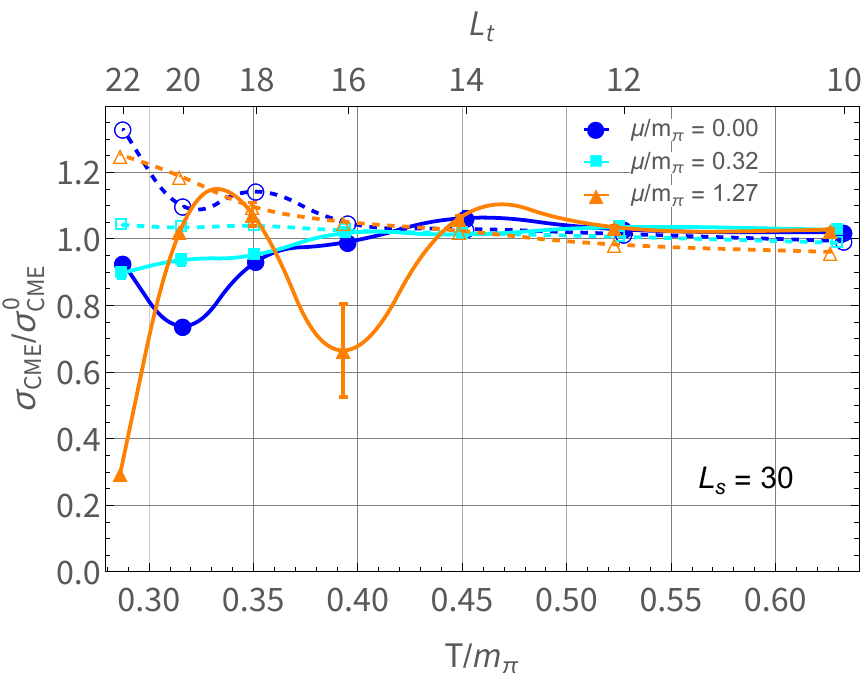}\\
	\caption{The ratio of the CME coefficient extracted from the axial-vector correlator (\ref{eq:CME_Euclidean_correlator}) to the free, continuum, chiral quark value $\sigma_{CME}^0 = \frac{e B \, N_c \, T}{2 \pi^2}$ as a function of temperature $T$ and fermion chemical potential $\mu$ for spatial lattice sizes $L_s = 24$ (left) and $L_s = 30$ (right). Data points with error bars joined by solid lines and empty plot markers joined by dashed lines correspond to the full gauge theory case and the free lattice quarks, respectively. Data points for different values of $\mu$ are slightly displaced along the horizontal axis to avoid overlap between different plot markers. Continuous lines drawn through data points with the same $\mu$ are quadratic splines intended to guide the eye.}
	\label{fig:CME_summary}
\end{figure*}

Fig.~\ref{fig:QAJV} suggests that the axial-vector correlator in $SU(2)$ gauge theory at finite density is also in good agreement with the free quark result (\ref{eq:CME_Euclidean_correlator}). While the signal-to-noise ratio gets worse towards larger densities and lower temperatures, the data is close to (but slightly exceeding) the free quark result (\ref{eq:CME_Euclidean_correlator}) for all values of $T$ and $\mu$ that we consider. We therefore characterize the strength $\sigma_{CME}$ of the real-time CME response in terms of the height of the plateau that we measure by averaging the axial-vector correlator between $\tau = 4 a$ and $\tau = \lr{L_t - 4} a$. In this interval, the axial-vector correlator is constant within statistical errors for all our values of $\mu$ and $T$. Its average value therefore estimates the plateau height and is not affected by contact terms at small $\tau$. In addition, we use two values $\Phi = 1, 2$ of the total flux $\Phi = \frac{e B a^2 L_s^2}{2 \pi}$ of external magnetic field through the lattice to demonstrate that the plateau height indeed scales linearly with $\Phi$ and hence with $B$. We then average the ratio of the plateau height to $\lr{e B} T C_{em}$ over these two values of the magnetic flux to arrive at a single estimate of the CME response at a given temperature and density. These estimates are shown on Fig.~\ref{fig:CME_summary} for two spatial lattice sizes, $L_s = 24$ and $L_s = 30$. Data points with error bars are connected using quadratic splines (solid lines) to guide the eye. For comparison, we also show the corresponding correlators for free Wilson-Dirac fermions with the same lattice parameters.

The plot suggests a reasonably good agreement with the free, continuum, massless quark result $\sigma_{CME}^0 = \frac{e \, B \, N_c \, T}{2 \pi^2}$ at almost all values of $T$ and $\mu$. The temperature dependence exhibits some downward oscillations at lower temperatures both for the full gauge theory and for free quarks, which become somewhat more pronounced towards larger chemical potential. Despite the oscillatory behavior, the CME coefficient obtained for the full gauge theory does not exceed the continuum, free-fermion value of $\sigma_{CME}^0$ by more than $10\%$.  

As already noticed in \cite{Buividovich:24:2}, lattice artifacts appear to be larger for free quarks than for the full gauge theory. In particular, at small temperatures the free quark result on the lattice deviates from the continuum value by as much as $50\%$. A comparison of the lattice results for free quarks on lattice sizes $L_s = 24$ and $L_s = 30$ suggests that these oscillations become less pronounced for $L_s = 30$, and are therefore likely to be finite-volume artifacts caused by subtle interplay between Landau levels, Fermi level, and finite volume \cite{Buividovich:13:7,Buividovich:13:8}. The fact that the full $SU\lr{2}$ gauge theory has smaller correlation length than free quarks is a viable explanation of less pronounced finite-volume artifacts. 

Overall, however, there are no signatures of any significant (e.g. order of magnitude) enhancement of the CME signal across the $\lr{\mu, T}$ plane, including the chiral crossover region at small $\mu \lesssim m_{\pi}/2$ and the second-order diquark condensation phase transition at larger $\mu$. Most of our data points at low temperatures are slightly below $\sigma_{CME}^0$ and also below the corresponding results for free lattice quarks, in qualitative agreement with mild CME suppression observed in finite-temperature, zero-density simulations of \cite{Brandt:2502.01155}.

\section{Negative magnetoresistance at finite fermion density}
\label{sec:NMR}

We now turn to the dependence of the electric conductivity on the external magnetic field which is widely considered to be a $\mathcal{CP}$-even proxy observable for the CME \cite{Fukushima:0912.2961,Buividovich:10:1,Braguta:2406.18504,Braguta:1910.08516,Kharzeev:1412.6543,Sun:1603.02624,Fukushima:2106.07968}. Needless to say, the electric conductivity of the quark-gluon plasma is an interesting physical property in itself, regardless of its relation to NMR and CME \cite{Greiner:1408.7049,Puglisi:1408.7043,Qin:1307.4587,Greiner:1602.05085,Fernandez:hep-ph/0512283}. In particular, its value affects the dilepton emission rate \cite{McLerran:85:1} and the lifetime of the strong magnetic fields generated in off-central heavy-ion collisions \cite{Skokov:1305.0774,Gursoy:2009.09727,Grieninger:2503.10593}.

The usual derivation leading to Negative Magnetoresistance (NMR) combines the paradigmatic expression for the CME current $\vec{j} = \frac{C_{em} \, N_c \, \mu_A \, \vec{B}}{2 \pi^2}$ with the real-time anomaly equation $\frac{d Q_A}{d t} = \frac{\vec{E} \cdot \vec{B}}{2 \pi^2} - \tau_A^{-1} \, Q_A$, where $\mu_A$ is the chiral chemical potential, $\tau_A$ is the phenomenological relaxation time of the axial charge, and $\vec{E}$ is the external electric field. One then assumes a linear relation $Q_A = \chi_A \, \mu_A$, where $\chi_A$ is the axial charge susceptibility, to arrive at the relation $\vec{j} \sim \frac{\tau_A}{\chi_A} \, \vec{B} \lr{\vec{B} \cdot \vec{E}}$. This immediately leads to the conclusion that the low-frequency limit of electric conductivity $\sigma_{\parallel}$ in the direction of the magnetic field, which we define as the corresponding diagonal component of the conductivity tensor $\sigma_{ik}$ (e.g. $\sigma_{zz}$ if $\vec{B} \parallel \vec{e}_z$), should scale quadratically with the magnetic field strength $\big| \vec{B}\big|$. Since the factor $\tau_A/\chi_A$ in general depends on temperature and density in a non-trivial way, the coefficient of the quadratic dependence of $\sigma_{\parallel}$ is not directly related to the CME response. Furthermore, $\sigma_{\parallel}$ in an interacting theory can also receive contributions that are not directly related to CME \cite{Kharzeev:1412.6543,Goswami:1503.02069}.

A more direct relation between $\sigma_{\parallel}$ and the CME can be established for free fermions within the linear response approximation \cite{Buividovich:24:2,Gorbar:1312.0027}. Namely, Green-Kubo formulas relate the AC electric-conductivity tensor $\sigma_{ik}\lr{\omega}$ to correlators of electric currents. In Euclidean time, this relation reads \cite{Meyer:1104.3708,Nikolaev:2008.12326}:
\begin{eqnarray}
\label{eq:GK_conductivity}
	\frac{1}{V} \int d^3\vec{x} \, \vev{j_i\lr{\tau, \vec{x}} j_k\lr{0, \vec{0}}}
	= \nonumber \\ =
	\int\limits_{0}^{\infty} d\omega  \, \frac{\omega}{\pi} \,
	\frac{\cosh\lr{\omega \lr{ \tau - \frac{1}{2T} }}}{\sinh\lr{\frac{\omega}{2 T}}} \, \sigma_{ik}\lr{\omega} ,
\end{eqnarray} 
For free quarks in external magnetic fields, the correlator (\ref{eq:GK_conductivity}) for longitudinal currents (currents in the direction of the magnetic field) can be decomposed into a sum of contributions from Landau levels labelled by integers $n = 0, 1, 2, \ldots, +\infty$. It turns out that the contribution of the lowest Landau level with $n=0$ to the longitudinal component of the correlator (\ref{eq:GK_conductivity}) is identically equal to the axial-vector correlator (\ref{eq:CME_Euclidean_correlator}) \cite{Gorbar:1312.0027,Buividovich:24:2}, and is hence directly related to the CME. An important difference between the longitudinal component of the vector-vector correlator (\ref{eq:GK_conductivity}) and the axial-vector correlator (\ref{eq:CME_Euclidean_correlator}) is that the contributions of all higher Landau levels with $n>0$ to the axial-vector correlator are zero \cite{Buividovich:24:2}. On the other hand, the vector-vector correlator in (\ref{eq:GK_conductivity}) receives contributions from all Landau levels with $n>0$, which hinder a direct extraction of the CME signal \cite{Gorbar:1312.0027}. We will see that detecting the CME signal on top of background contributions becomes even harder in the full $SU(2)$ gauge theory at finite $T$ and $\mu$.

Finite chemical potential is known to lead to a strong enhancement of the electric conductivity \cite{Cassing:1312.3189,Bratkovskaya:1911.08547,Smekal:1807.04952,Kadam:1712.03805,Ghosh:1607.01340,Buividovich:20:1}, which is not surprising given that quarks are electric-charge carriers, and by increasing the chemical potential we increase the charge-carrier concentration. The electric conductivity of finite-temperature, finite-density $SU(2)$ gauge theory in the absence of external magnetic fields was studied in detail previously in \cite{Buividovich:20:1}.

\begin{figure*}[h!tpb]
	\includegraphics[width=0.49\textwidth]{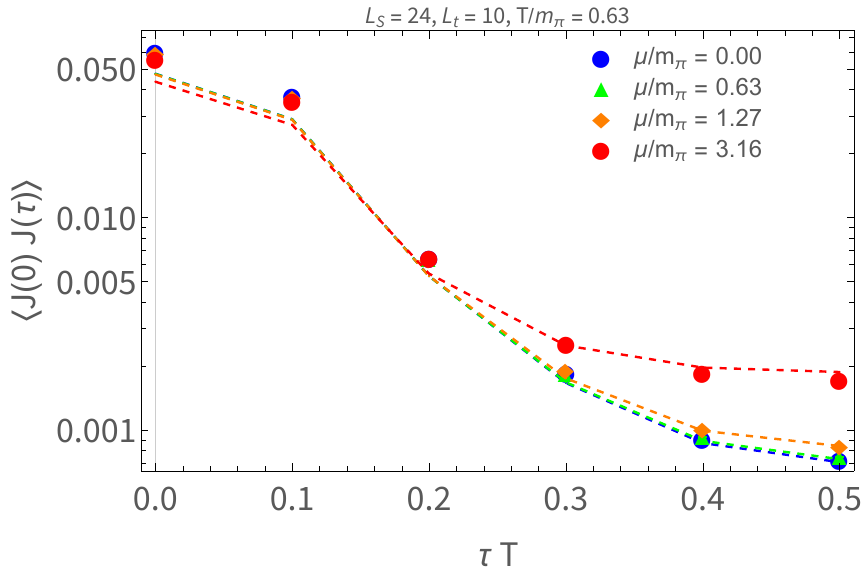}
	\includegraphics[width=0.49\textwidth]{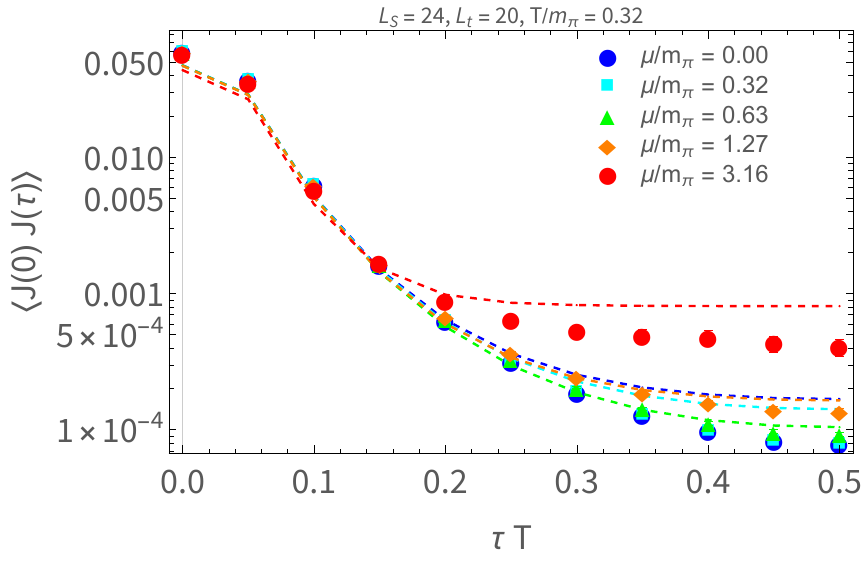}\\
	\caption{Correlators of vector currents as functions of Euclidean time $\tau$ at zero external magnetic field and at different values of temperature and fermion chemical potential. $T/m_{\pi} = 0.63$ ($L_t = 10$) on the left, and $T/m_{\pi} = 0.32$ ($L_t = 20$) on the right. Lattice size is $L_s = 24$ in lattice units. Data points with error bars and dashed lines correspond to the full gauge theory case and the free lattice quarks, respectively.}
    \label{fig:JVJVLT}
\end{figure*}

\begin{figure*}[h!tpb]
	\includegraphics[width=0.49\textwidth]{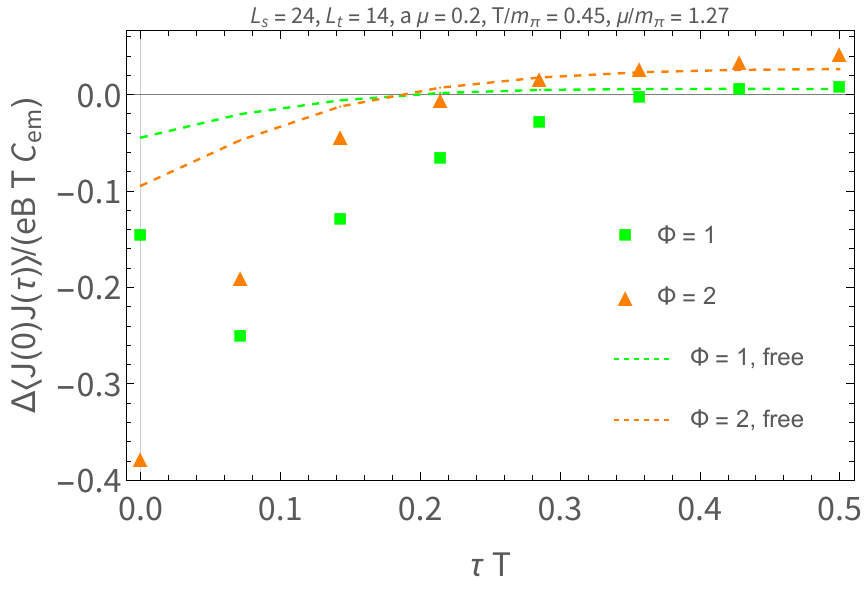}
	\includegraphics[width=0.49\textwidth]{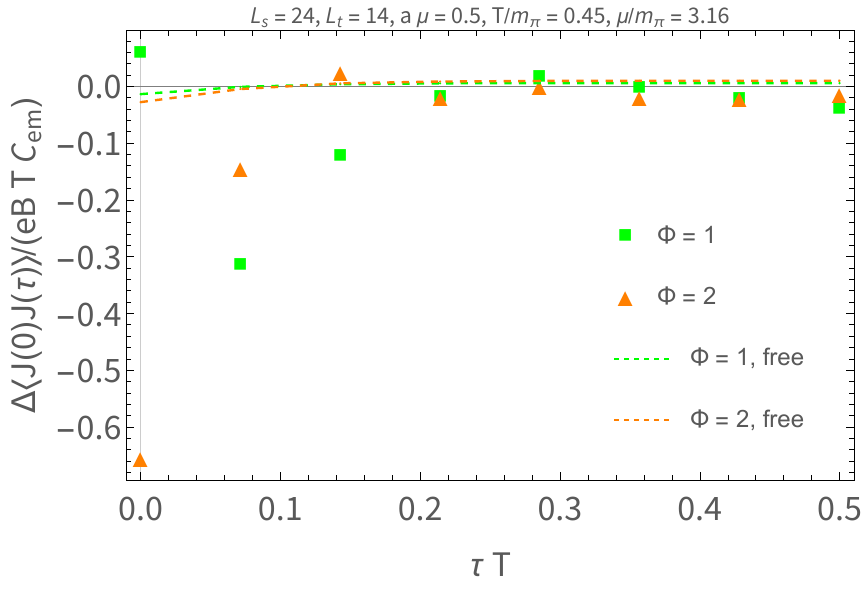}
	\includegraphics[width=0.49\textwidth]{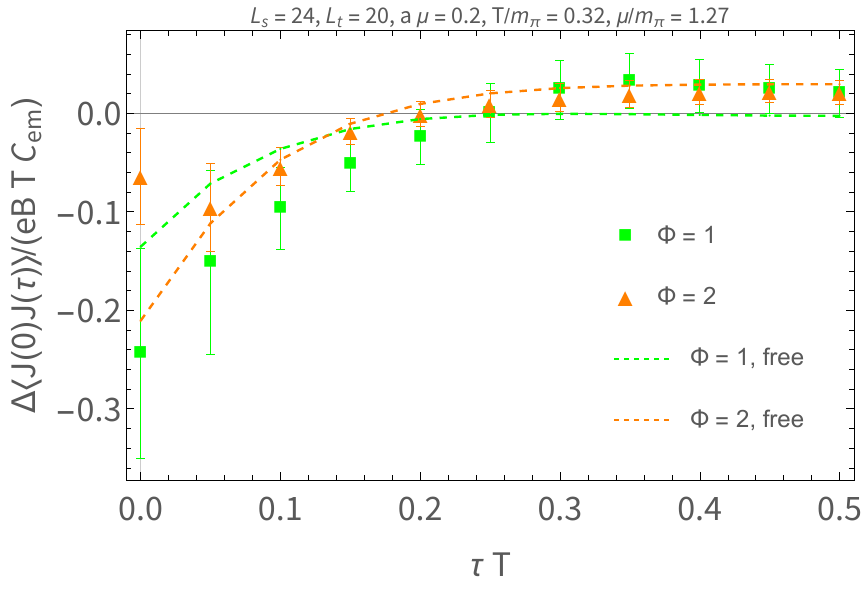}
	\includegraphics[width=0.49\textwidth]{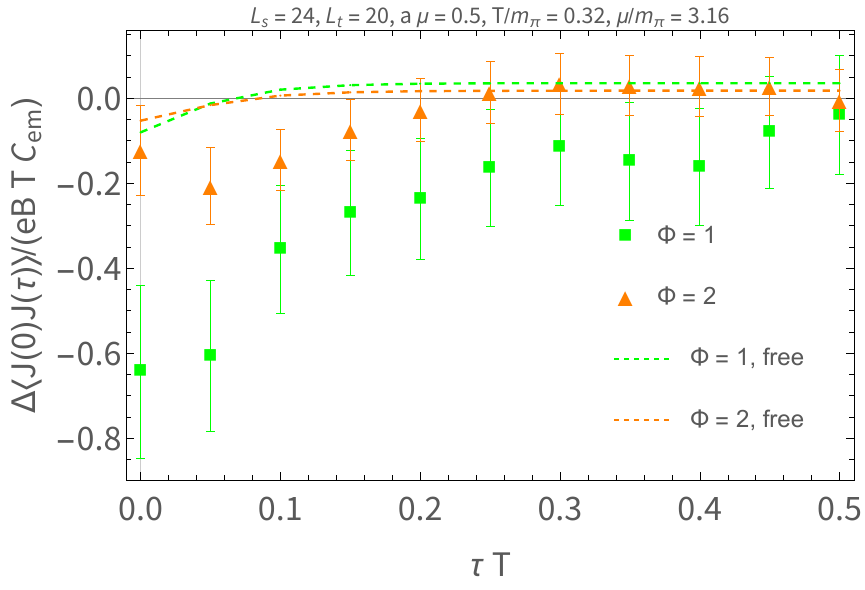}
	\caption{Subtracted correlators of vector currents as functions of Euclidean time $\tau$ at different values of temperature, chemical potential, and magnetic field strength $e B = \frac{2 \pi \Phi}{L_s^2 a^2}$, where $\Phi$ is the total magnetic flux. Lattice size is $L_s = 24$ in lattice units. Data points with error bars and dashed lines correspond to the full gauge theory case and the free lattice quarks, respectively.}
    \label{fig:DJVJV}
\end{figure*}

For reference, in Fig.~\ref{fig:JVJVLT} we show the corresponding current-current correlators (\ref{eq:GK_conductivity}) in absence of an external magnetic field on the lattice with spatial size $L_s = 24$. At both, low temperatures (right plot for $L_t = 20$) and high temperatures (left plot for $L_t = 10$), the current-current correlators become larger and shallower towards higher densities, indicating an increase in the low-frequency limit of the electric conductivity. The effect of an external magnetic field on these correlators is much smaller in scale, and to make it visible, in Fig.~\ref{fig:DJVJV} we plot the current-current correlators (\ref{eq:GK_conductivity}) at non-zero magnetic field with the zero-field results subtracted, 
\begin{eqnarray}
	V^{-1} \int d^3\vec{x} \, \left( \vev{j_i\lr{\tau, \vec{x}} j_k\lr{0, \vec{0}}}_{\vec{B}} 
- \right. \nonumber \\ \left. - 
\vev{j_i\lr{\tau, \vec{x}} j_k\lr{0, \vec{0}}}_{\vec{B}=0} \right). 
\nonumber
\end{eqnarray}
As in the previous section, we concentrate on the regime of small magnetic fields and use only the smallest non-zero values of the total magnetic flux, $\Phi = 1$ and $\Phi = 2$. To facilitate comparison with the CME analysis in the previous section, the subtracted vector-vector correlators in Fig.~\ref{fig:DJVJV} are normalized by $e B \, T \, C_{em}$ in the same way as the axial-vector correlators shown in Fig.~\ref{fig:QAJV}. The latter approach a plateau of height $\frac{N_c \, e B \, T \, C_{em}}{2 \pi^2} \approx 0.101 \, \lr{e B \, T \, C_{em}}$ at intermediate $\tau$. This normalization is motivated by the similarity between the lowest Landau-level contributions to the axial-vector and vector-vector correlators (\ref{eq:CME_Euclidean_correlator}) and (\ref{eq:GK_conductivity}), characterizing the CME and NMR, respectively, in the free massless continuum quark limit. The data in Fig.~\ref{fig:DJVJV} indeed indicates that for small and moderate densities magnetic fields cause some statistically significant increase in the vector-vector correlator at intermediate $\tau$. However, the scale of this increase is considerably smaller than the lowest Landau level contribution $\frac{N_c \, e B \, T \, C_{em}}{2 \pi^2}$ in the axial-vector correlator. For larger values of the chemical potential, the external magnetic fields hardly affect the current-current correlator beyond statistical errors. We also remark that, as usual for Euclidean-time correlators, the data are strongly correlated across different values of Euclidean time. The error bars shown in Fig.~\ref{fig:DJVJV} correspond to the diagonal entries of the full covariance matrix and only serve as a rough qualitative estimate of statistical errors. The full covariance matrix is used, however, to extract the electric conductivity from these data.

The relatively weak dependence of the electric conductivity on external magnetic fields $e B \lesssim m_{\pi}^2$ (as compared to its dependence on the chemical potential) makes the application of conventional methods for the extraction of the electric conductivity $\sigma\lr{\omega}$ from the Euclidean-time correlator (\ref{eq:GK_conductivity}) impractical. For example, with the regularized Backus-Gilbert method \cite{Meyer:1104.3708,Ulybyshev:1707.04212,Buividovich:20:1}, the $\tau$-dependent weights of $\vev{j_i\lr{\tau, \vec{x}} j_k\lr{0, \vec{0}}}$ in the conductivity estimates become large and sign-alternating at small values of the Tikhonov regularization parameter. As a result, the conductivity value depends on delicate cancellations between statistically uncertain correlator values at different $\tau$, and the statistical errors are quickly inflated. This makes it impossible to keep the errors sufficiently small while minimizing the frequency resolution. In this situation, we use a simpler and numerically much more stable mid-point estimator \cite{Kaczmarek:1012.4963}. Namely, according to the Green-Kubo relation in (\ref{eq:GK_conductivity}), the current-current correlator at the maximal Euclidean time separation $\tau = \frac{1}{2 T}$ is related to the AC electric conductivity $\sigma_{ik}\lr{\omega}$ as
\begin{eqnarray}
\label{eq:GK_MP1}
	\frac{1}{V} \int d^3\vec{x} \, \vev{j_i\lr{\frac{1}{2 T}, \vec{x}} j_k\lr{0, \vec{0}}}
	= \nonumber \\ =
	\int\limits_{0}^{\infty} d\omega  \, 
	\frac{\omega}{\pi \, \sinh\lr{\frac{\omega}{2 T}}} \, \sigma_{ik}\lr{\omega} .
\end{eqnarray}
The function $\frac{\omega}{\pi \, \sinh\lr{\frac{\omega}{2 T}}}$ is localized within the region of small frequencies $\omega \sim T$ and can be also considered as a ``smeared'' $\delta$-function similar to the one used in the Backus-Gilbert method. The norm and width of this function are:
\begin{eqnarray}
\label{eq:GK_MP2}
	\mathcal{N} \equiv \int\limits_{0}^{\infty} d \omega
	\, \frac{\omega}{\pi \, \sinh\lr{\frac{\omega}{2 T}}} = \pi T^2 ,
	\nonumber \\
	\Delta\omega = \sqrt{\mathcal{N}^{-1} \, \int\limits_{0}^{\infty} d\omega\, \omega^2 \, \frac{\omega}{\pi \, \sinh\lr{\frac{\omega}{2 T}}} }
	= \nonumber \\ =
	\sqrt{2} \pi T \approx 4.4 \, T.
\end{eqnarray}
We can thus use the value of the Euclidean correlator (\ref{eq:GK_conductivity}) at the midpoint $\tau = \frac{1}{2 T}$ as an estimate of the electric conductivity $\sigma_{ik}\lr{\omega}$ smeared over frequencies in the range $\omega \lesssim 4.4 \, T$:
\begin{eqnarray}
\label{eq:GK_MP3}
	\sigma_{ik} \approx \frac{1}{\pi T^2}\, \frac{1}{V} \int d^3\vec{x} \, \vev{j_i\lr{\frac{1}{2 T}, \vec{x}} j_k\lr{0, \vec{0}}} .
\end{eqnarray}
In contrast to Backus-Gilbert estimators, the mid-point estimator does not include large opposite-sign weights and has the same statistical errors as the original current-current correlator.

\begin{figure*}[h!tpb]
	\includegraphics[width=0.49\textwidth]{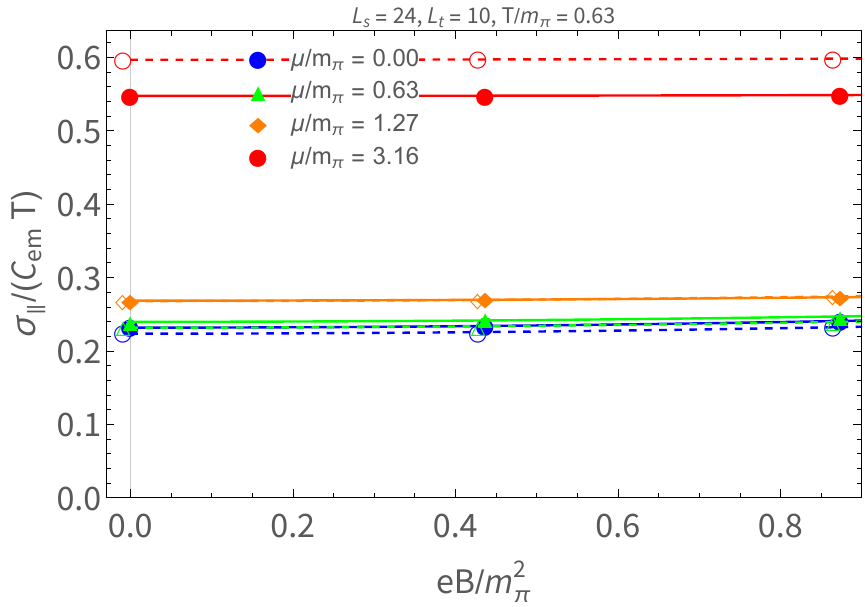}
	\includegraphics[width=0.49\textwidth]{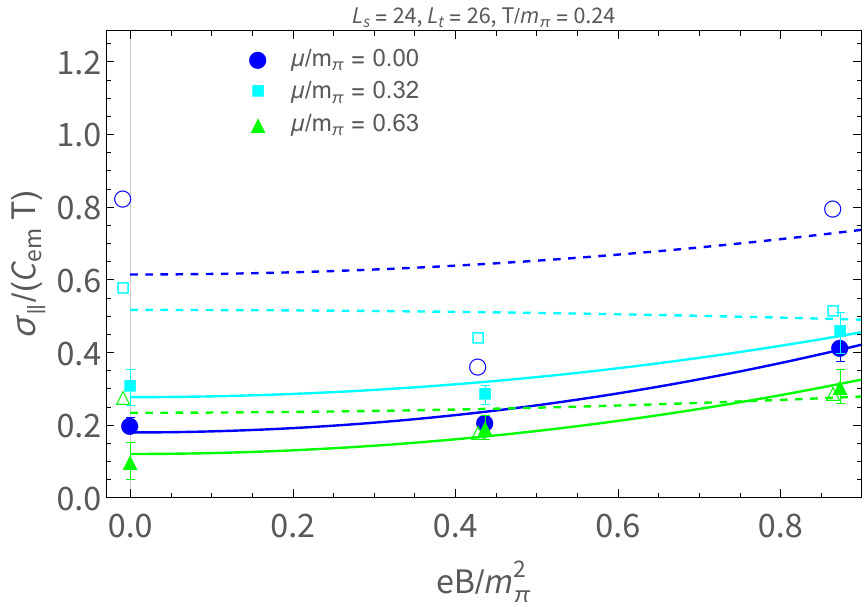}
	\caption{Mid-point estimates (\ref{eq:GK_MP3}) of the longitudinal component of the low-frequency limit of the electric conductivity $\sigma_{\parallel}$ as a function of the magnetic field $B$ for different values of temperature and fermion chemical potential. Solid lines represent quadratic fits of the form (\ref{eq:nmr_def}). Empty plot markers correspond to the results obtained for free quarks on the lattice with the same parameters as in the full gauge theory, and dashed lines are the fits of the same form (\ref{eq:nmr_def}) to the free quark data. Data points for different values of $\mu$ are slightly displaced along the horizontal axis to avoid overlap between different plot markers. }
	\label{fig:NMR_examples}
\end{figure*}

For completeness, we also analyse all of our data using the Tikhonov-regularized Backus-Gilbert method \cite{Ulybyshev:1707.04212} with the same values of regularization parameter as in \cite{Buividovich:20:1}. This yields the results that are very similar to the ones obtained from the mid-point estimator (\ref{eq:GK_MP3}), but with considerably larger statistical errors. 

\begin{figure*}[h!tpb]
	\includegraphics[width=0.47\textwidth]{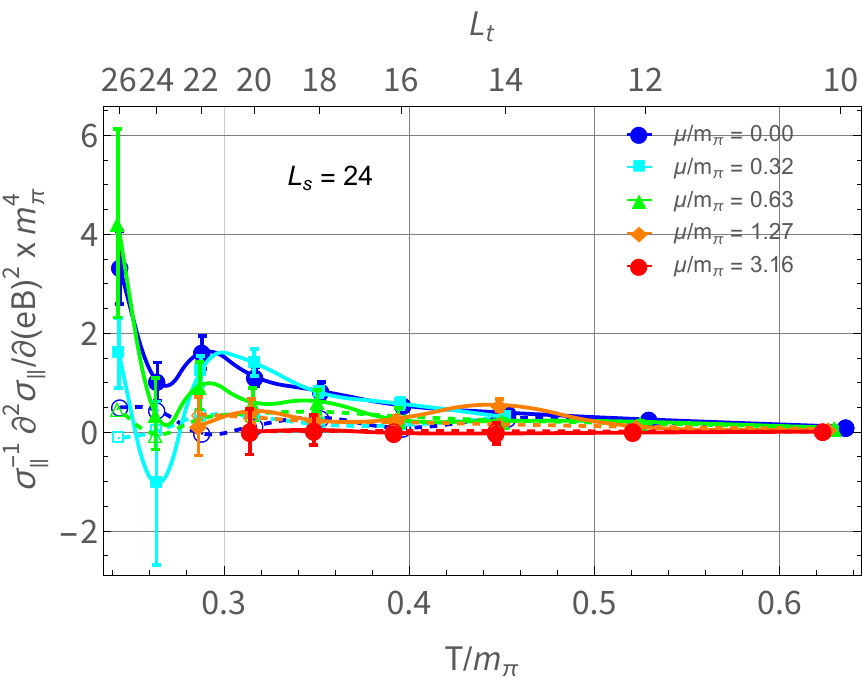}{ }\includegraphics[width=0.47\textwidth]{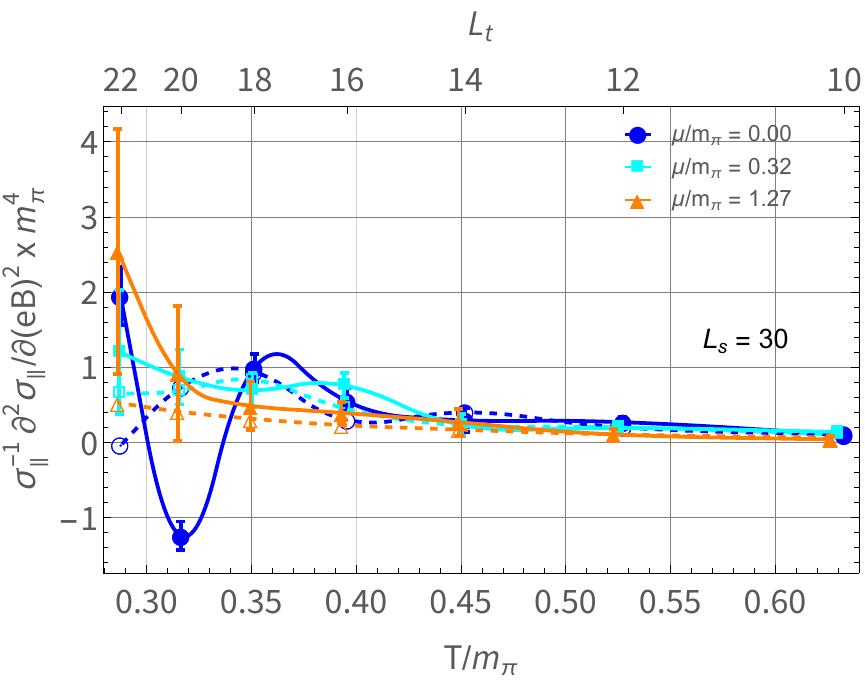}\\
	\caption{NMR coefficient $\sigma_{\parallel}^{-1} \, \frac{\partial^2 \sigma_{\parallel}}{\partial \lr{e B}^2}$ estimated from the fit of the form (\ref{eq:nmr_def}) as a function of temperature $T$ and fermion chemical potential $\mu$ for spatial lattice sizes $L_s = 24$ (left) and $L_s = 30$ (right). We use the mid-point estimator (\ref{eq:GK_MP3}) for the low-frequency limit $\sigma_{\parallel}\lr{\omega \rightarrow 0}$ of the longitudinal electric conductivity. Data points with error bars joined by solid lines and empty plot markers joined by dashed lines correspond to the full gauge-theory case and free lattice quarks, respectively. Data points for different values of $\mu$ are slightly displaced along the horizontal axis to avoid overlap between different plot markers. Continuous lines drawn through data points with the same $\mu$ are quadratic splines intended to guide the eye.}
	\label{fig:NMR_summary}
\end{figure*}
 
The mid-point estimates of the electric conductivity as a function of the external magnetic field-strength $e B$ are shown in Fig.~\ref{fig:NMR_examples} for different values of chemical potential and temperature. These plots clearly show that the dependence of the conductivity on the magnetic field for $e B \lesssim m_{\pi}^2$ is much weaker than its density dependence. Solid continuous lines in Fig.~\ref{fig:NMR_examples} show single-parameter fits of the form 
\begin{eqnarray}
\label{eq:nmr_def}
	\frac{\sigma_{\parallel}\lr{T, \mu, B}}{T \, C_{em}} = \frac{\sigma_{\parallel}\lr{T, \mu, 0}}{T \, C_{em}} \, \lr{ 1 + \frac{c\lr{T, \mu} \, \lr{e B}^2}{2} } ,
\end{eqnarray}
where the only fit parameter $c\lr{T, \mu}$ can be identified with the second derivative $\sigma_{\parallel}^{-1} \frac{\partial^2 \sigma_{\parallel}}{\partial \lr{e B}^2}$. A closer look at the fits in Fig.~\ref{fig:NMR_examples} suggests that the quadratic fits (\ref{eq:nmr_def}) may not be perfect, but we found that linear fits of the form $\sigma_{\parallel}\lr{T, \mu, B} = \sigma_{\parallel}\lr{T, \mu} \, \lr{1 + c \, \abs{B}}$, which mimic the lower Landau level contribution to $\sigma_{\parallel}$,  give a slightly worse description of the data, in general. Guided by the conventional NMR derivation in terms of CME as well as the more advanced analysis of \cite{Fukushima:1906.02683,Shaikh:2210.15388,Thakur:1910.12087} and numerical results of \cite{Braguta:1707.09810}, we therefore use the quadratic fit (\ref{eq:nmr_def}) to estimate the second derivative $\sigma_{\parallel}^{-1} \frac{\partial^2 \sigma_{\parallel}}{\partial \lr{e B}^2}$ and characterize the Negative Magnetoresistance for small magnetic fields $e B \lesssim m_{\pi}^2$. According to most numerical estimates \cite{Skokov:0907.1396,Tuchin:1305.5806,Gursoy:2009.09727,Grieninger:2503.10593}, such field strengths are more relevant for off-central heavy-ion collisions at RHIC than asymptotically large magnetic fields with $e B \gg m_{\pi}^2$.

\begin{figure}[h!tpb]
	\includegraphics[width=0.49\textwidth]{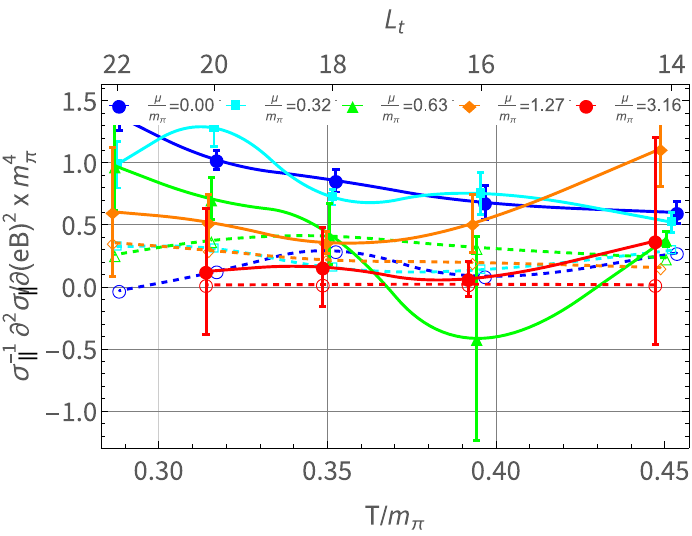}
	\caption{NMR coefficient $\sigma_{\parallel}^{-1} \frac{\partial^2 \sigma_{\parallel}}{\partial \lr{e B}^2}$ extracted from the quadratic fit (\ref{eq:nmr_def}) as a function of temperature $T$ and fermion chemical potential $\mu$ for spatial lattice size $L_s = 24$. Here we use the Tikhonov-regularized Backus-Gilbert method \cite{Ulybyshev:1707.04212,Buividovich:20:1} to estimate the low-frequency limit $\sigma_{\parallel}\lr{w \rightarrow 0}$ of longitudinal electric conductivity. Data points with error bars joined by solid lines and empty plot markers joined by dashed lines correspond to the full gauge theory case and the free lattice quarks, respectively. Data points for different values of $\mu$ are slightly displaced along the horizontal axis to avoid overlap between different plot markers. Continuous lines drawn through data points with the same $\mu$ are quadratic splines that are intended to guide the eye.}
	\label{fig:NMR_summary_BG}
\end{figure}

Our final estimates of $\sigma_{\parallel}^{-1} \frac{\partial^2 \sigma_{\parallel}}{\partial \lr{e B}^2}$ obtained from the mid-point value of current-current correlators are summarized in Fig.~\ref{fig:NMR_summary} for our two spatial lattice sizes $L_s = 24$ and $L_s = 30$. Data points with error bars are connected using quadratic splines (solid lines) to guide the eye. In qualitative agreement with previous numerical results \cite{Buividovich:10:1,Braguta:1910.08516,Braguta:2406.18504}, we find that at zero chemical potential Negative Magnetoresistance becomes more pronounced towards lower temperatures, where it also deviates significantly from the free quark result (dashed lines in Fig.~\ref{fig:NMR_summary}). Finite density tends to dampen the NMR response overall in comparison with the zero-density case, except for a couple of data points that are only slightly above the zero-density result. We note that one of these data points ($\mu/m_{\pi} = 0.63$, $T/m_{\pi} = 0.24$) corresponds to the vicinity of the intersection of the chiral crossover and diquark condensation transition in Fig.~\ref{fig:phase_diagram}. While our data does not exclude a mild enhancement of NMR in the vicinity of a 2nd-order diquark condensation transition at low temperatures, the difference between the results at zero and non-zero densities is not statistically significant for our data. At larger values of the chemical potential, $\mu \gtrsim m_{\pi}$, the NMR response becomes noticeably suppressed for all temperatures that we consider, including the vicinity of a 2nd order diquark condensation transition.

The results obtained with the two different lattice sizes $L_s = 24$ and $L_s = 30$ are consistent with each other within statistical errors. Again, we observe that finite-volume artifacts appear to be larger for free quarks than for the full gauge theory.

An interesting feature of the NMR response that appears at finite densities is the non-monotonous dependence of $\sigma_{\parallel}^{-1} \frac{\partial^2 \sigma_{\parallel}}{\partial \lr{e B}^2}$ on $\mu$ and $T$ at $\mu \lesssim m_{\pi}/2$ and $T \lesssim T_c$. While it might be tempting to associate this behavior with the vicinity of the chiral crossover, we note that the free-fermion results exhibit qualitatively similar non-monotonicity. Given that we effectively estimate the second derivative over $e B$ in terms of finite differences for fluxes $\Phi = 1$ and $\Phi = 2$, an interplay between Landau levels, Fermi level, and discrete lattice momenta (akin to Shubnikov-de Haas oscillations) might be a viable explanation for this non-monotonic behavior. A detailed investigation of this phenomenon goes beyond the scope of this paper.

Finally, for completeness, in Fig.~\ref{fig:NMR_summary_BG} we also present the estimates of the NMR coefficient $\sigma_{\parallel}^{-1} \frac{\partial^2 \sigma_{\parallel}}{\partial \lr{e B}^2}$ with $\sigma_{\parallel}$ obtained from the Tikhonov-regularized Backus-Gilbert method \cite{Ulybyshev:1707.04212,Buividovich:20:1}. The values of the regularization parameter are identical to those used in our work \cite{Buividovich:20:1}. Overall, the results obtained using the Backus-Gilbert method and the mid-point estimator (\ref{eq:nmr_def}) (Fig.~\ref{fig:NMR_summary}) are in good agreement, but statistical uncertainties are much larger in the case of the Backus-Gilbert method.

\section{Conclusions and discussion}
\label{sec:conclusions}

In this paper, we studied two physical observables related to the Chiral Magnetic Effect (CME) across the phase diagram of $SU(2)$ gauge theory with $N_f = 2$ flavours of light dynamical quarks at finite chemical potential $\mu$ and temperature $T$. Our analysis is limited to the smallest possible non-zero values of the magnetic field on finite-volume lattices. On the one hand, this restricts our analysis to relatively small magnetic field strengths $e B \lesssim m_{\pi}^2$, which are relevant for Beam Energy Scan II (as well as most other experiments) at RHIC. On the other hand, the smallness of the external magnetic field provides a rationale for our ``magnetic-quenched'' approximation with the magnetic field affecting only the valence quarks. This approximation is necessary to ensure the absence of the fermionic sign problem in $SU(2)$  gauge theory that would otherwise occur when both, chemical potential and magnetic field are non-zero.

The first observable is a $\mathcal{CP}$-odd axial-vector correlator in an external magnetic field, which gives direct access to the out-of-equilibrium CME electric current generated by thermal fluctuations of the axial charge density \cite{Buividovich:24:2}. Our overall conclusion is that this observable exhibits a rather weak density dependence, and does not deviate significantly from the corresponding free-quark results (\ref{eq:CME_Euclidean_correlator}). There are some indications of a weak suppression in the hadronic phase at low temperatures and large densities. Calculations in $SU(3)$ gauge theory suggest that this suppression might become quite significant at very low $T$ \cite{Brandt:2502.01155}. In the high-temperature quark-gluon plasma regime, the effect of finite density on the CME seems to be very small.

Another observable is a $\mathcal{CP}$-even correlator of electric (vector) currents, which is related to the electric conductivity by virtue of Green-Kubo relations. Negative Magnetoresistance (NMR), a growth of longitudinal electric conductivity $\sigma_{\parallel}$ with magnetic field, is often considered as one of the main signatures of the CME \cite{Fukushima:0912.2961,Buividovich:10:1,Braguta:2406.18504,Braguta:1910.08516,Kharzeev:1412.6543,Sun:1603.02624,Fukushima:2106.07968,Braguta:1707.09810}. The relation between vector-vector and axial-vector correlators can be made very precise for free massless quarks, where they receive identical contributions from the lowest Landau level.

We found that compared to the axial-vector correlator, the vector-vector correlator and hence the longitudinal electric conductivity $\sigma_{\parallel}$ have a much weaker dependence on the strength of the external magnetic field. The absolute increase in $\sigma_{\parallel}$ due to the external field is considerably smaller than the lowest Landau level contribution, for all values of $T$ and $\mu$ that we considered. Furthermore, the CME and NMR seem to exhibit a qualitatively different behaviour in the $\lr{T, \mu}$ plane: while the CME appears to be suppressed at low temperatures \cite{Brandt:2502.01155}, the NMR is most pronounced in the hadronic regime at low temperatures and small densities. NMR is strongly suppressed towards large densities, where CME also exhibits an oscillatory downward trend. However, the density dependence of NMR is much stronger.

This comparison between CME and NMR observables suggests that at least in the weak-field regime (most relevant for RHIC) NMR might not be an optimal probe of the CME. It would therefore be advantageous to construct refined experimental observables,   sensitive to axial-vector rather than vector-vector correlations.

Our results also do not provide any statistically significant evidence of a strong enhancement of either the CME or NMR responses in the vicinity of a chiral crossover or the second-order phase transition line in the $\lr{\mu, T}$ phase diagram. The only part of the phase diagram where we cannot exclude a mild enhancement of NMR is the vicinity of a 2nd-order diquark condensation line at low temperatures and for chemical potentials close to diquark condensation threshold at $\mu = m_{\pi}/2$. A conclusive first-principles study of this regime should involve simulations at low temperatures and several sufficiently large lattice volumes, and would therefore be computationally very demanding. Since our current gauge ensembles do not allow to make statistically significant conclusions in this particular regime, we leave such a study for future work.

Of course, these results might well be specific to the $SU(2)$ gauge theory considered here, where qualitative similarity of the phase diagram and physical observables to real QCD can only be expected at $\mu \lesssim m_{\pi}/2$, or temperatures and densities outside the diquark condensation phase. If this is the case, the mechanism of possible CME enhancement in the recent results of the STAR collaboration \cite{STAR:2506.00275} might well be non-universal, or driven by factors other than temperature and chemical potential alone. 

\begin{acknowledgments}
The work of P.B. was funded in part by the UK STFC Consolidated Grant ST/X000699/1. Numerical simulations were undertaken on Barkla, part of the High Performance Computing facilities at the University of Liverpool, UK, and on the GPU cluster of the Institute for Theoretical Physics at JLU Giessen. The authors are grateful to D.~Kharzeev for interesting discussions that motivated this work, and to the organizers of the workshop \href{https://indico.lip.pt/event/1991/}{``Weyl and Dirac Semimetals as a Laboratory for High-Energy Physics''} (Braga, June 25-28, 2025) for creating a stimulating discussion environment.
\end{acknowledgments}
	
%\bibliographystyle{apsrev4-2}
%\bibliography{Buividovich,CME_phase_diagram}

\begin{thebibliography}{92}%
\makeatletter
\providecommand \@ifxundefined [1]{%
 \@ifx{#1\undefined}
}%
\providecommand \@ifnum [1]{%
 \ifnum #1\expandafter \@firstoftwo
 \else \expandafter \@secondoftwo
 \fi
}%
\providecommand \@ifx [1]{%
 \ifx #1\expandafter \@firstoftwo
 \else \expandafter \@secondoftwo
 \fi
}%
\providecommand \natexlab [1]{#1}%
\providecommand \enquote  [1]{``#1''}%
\providecommand \bibnamefont  [1]{#1}%
\providecommand \bibfnamefont [1]{#1}%
\providecommand \citenamefont [1]{#1}%
\providecommand \href@noop [0]{\@secondoftwo}%
\providecommand \href [0]{\begingroup \@sanitize@url \@href}%
\providecommand \@href[1]{\@@startlink{#1}\@@href}%
\providecommand \@@href[1]{\endgroup#1\@@endlink}%
\providecommand \@sanitize@url [0]{\catcode `\\12\catcode `\$12\catcode `\&12\catcode `\#12\catcode `\^12\catcode `\_12\catcode `\%12\relax}%
\providecommand \@@startlink[1]{}%
\providecommand \@@endlink[0]{}%
\providecommand \url  [0]{\begingroup\@sanitize@url \@url }%
\providecommand \@url [1]{\endgroup\@href {#1}{\urlprefix }}%
\providecommand \urlprefix  [0]{URL }%
\providecommand \Eprint [0]{\href }%
\providecommand \doibase [0]{https://doi.org/}%
\providecommand \selectlanguage [0]{\@gobble}%
\providecommand \bibinfo  [0]{\@secondoftwo}%
\providecommand \bibfield  [0]{\@secondoftwo}%
\providecommand \translation [1]{[#1]}%
\providecommand \BibitemOpen [0]{}%
\providecommand \bibitemStop [0]{}%
\providecommand \bibitemNoStop [0]{.\EOS\space}%
\providecommand \EOS [0]{\spacefactor3000\relax}%
\providecommand \BibitemShut  [1]{\csname bibitem#1\endcsname}%
\let\auto@bib@innerbib\@empty
%</preamble>
\bibitem [{\citenamefont {{M.~S.~Abdallah~et~al. (STAR collaboration)}}(2021)}]{STAR:2109.00131}%
  \BibitemOpen
  \bibfield  {author} {\bibinfo {author} {\bibnamefont {{M.~S.~Abdallah~et~al. (STAR collaboration)}}},\ }\href {http://dx.doi.org/10.1103/PhysRevC.105.014901} {\bibfield  {journal} {\bibinfo  {journal} {Phys.~Rev.~C}\ }\textbf {\bibinfo {volume} {105}},\ \bibinfo {pages} {014901} (\bibinfo {year} {2021})},\ \Eprint {https://arxiv.org/abs/2109.00131} {2109.00131} \BibitemShut {NoStop}%
\bibitem [{\citenamefont {Aboona}\ \emph {et~al.}(2023)\citenamefont {Aboona} \emph {et~al.}}]{STAR:2209.03467}%
  \BibitemOpen
  \bibfield  {author} {\bibinfo {author} {\bibfnamefont {B.}~\bibnamefont {Aboona}} \emph {et~al.} (\bibinfo {collaboration} {STAR}),\ }\href {https://doi.org/10.1016/j.physletb.2023.137779} {\bibfield  {journal} {\bibinfo  {journal} {Phys. Lett. B}\ }\textbf {\bibinfo {volume} {839}},\ \bibinfo {pages} {137779} (\bibinfo {year} {2023})},\ \Eprint {https://arxiv.org/abs/2209.03467} {arXiv:2209.03467 [nucl-ex]} \BibitemShut {NoStop}%
\bibitem [{\citenamefont {Kharzeev}\ \emph {et~al.}(2022)\citenamefont {Kharzeev}, \citenamefont {Liao},\ and\ \citenamefont {Shi}}]{Kharzeev:2022hqz}%
  \BibitemOpen
  \bibfield  {author} {\bibinfo {author} {\bibfnamefont {D.~E.}\ \bibnamefont {Kharzeev}}, \bibinfo {author} {\bibfnamefont {J.}~\bibnamefont {Liao}},\ and\ \bibinfo {author} {\bibfnamefont {S.}~\bibnamefont {Shi}},\ }\href {https://doi.org/10.1103/PhysRevC.106.L051903} {\bibfield  {journal} {\bibinfo  {journal} {Phys. Rev. C}\ }\textbf {\bibinfo {volume} {106}},\ \bibinfo {pages} {L051903} (\bibinfo {year} {2022})},\ \Eprint {https://arxiv.org/abs/2205.00120} {arXiv:2205.00120 [nucl-th]} \BibitemShut {NoStop}%
\bibitem [{\citenamefont {Lacey}\ and\ \citenamefont {Magdy}(2022)}]{Lacey:2022plw}%
  \BibitemOpen
  \bibfield  {author} {\bibinfo {author} {\bibfnamefont {R.~A.}\ \bibnamefont {Lacey}}\ and\ \bibinfo {author} {\bibfnamefont {N.}~\bibnamefont {Magdy}},\ }\href@noop {} {\bibinfo {title} {{Scaling properties of the $\Delta\gamma$ correlator and their implication for detection of the chiral magnetic effect in heavy-ion collisions}}} (\bibinfo {year} {2022}),\ \Eprint {https://arxiv.org/abs/2206.05773} {arXiv:2206.05773} \BibitemShut {NoStop}%
\bibitem [{\citenamefont {Li}\ \emph {et~al.}(2025)\citenamefont {Li}, \citenamefont {Feng},\ and\ \citenamefont {Wang}}]{Li:2024gdz}%
  \BibitemOpen
  \bibfield  {author} {\bibinfo {author} {\bibfnamefont {H.-S.}\ \bibnamefont {Li}}, \bibinfo {author} {\bibfnamefont {Y.}~\bibnamefont {Feng}},\ and\ \bibinfo {author} {\bibfnamefont {F.}~\bibnamefont {Wang}},\ }\href {https://doi.org/10.1103/kwtd-bldv} {\bibfield  {journal} {\bibinfo  {journal} {Phys. Rev. C}\ }\textbf {\bibinfo {volume} {112}},\ \bibinfo {pages} {054901} (\bibinfo {year} {2025})},\ \Eprint {https://arxiv.org/abs/2407.14489} {arXiv:2407.14489 [physics.data-an]} \BibitemShut {NoStop}%
\bibitem [{\citenamefont {Guo}\ \emph {et~al.}(2025)\citenamefont {Guo}, \citenamefont {Wang}, \citenamefont {Zhou},\ and\ \citenamefont {Ma}}]{Guo:2025wry}%
  \BibitemOpen
  \bibfield  {author} {\bibinfo {author} {\bibfnamefont {S.}~\bibnamefont {Guo}}, \bibinfo {author} {\bibfnamefont {L.}~\bibnamefont {Wang}}, \bibinfo {author} {\bibfnamefont {K.}~\bibnamefont {Zhou}},\ and\ \bibinfo {author} {\bibfnamefont {G.-L.}\ \bibnamefont {Ma}},\ }\href {https://doi.org/10.1088/0256-307X/42/11/110101} {\bibfield  {journal} {\bibinfo  {journal} {Chin. Phys. Lett.}\ }\textbf {\bibinfo {volume} {42}},\ \bibinfo {pages} {110101} (\bibinfo {year} {2025})},\ \Eprint {https://arxiv.org/abs/2507.05808} {arXiv:2507.05808 [nucl-th]} \BibitemShut {NoStop}%
\bibitem [{\citenamefont {Feng}\ \emph {et~al.}(2025)\citenamefont {Feng}, \citenamefont {Voloshin},\ and\ \citenamefont {Wang}}]{Feng:2502.09742}%
  \BibitemOpen
  \bibfield  {author} {\bibinfo {author} {\bibfnamefont {Y.}~\bibnamefont {Feng}}, \bibinfo {author} {\bibfnamefont {S.~A.}\ \bibnamefont {Voloshin}},\ and\ \bibinfo {author} {\bibfnamefont {F.}~\bibnamefont {Wang}},\ }\href {https://doi.org/10.1103/xk4f-wm9s} {\bibfield  {journal} {\bibinfo  {journal} {Phys. Rev. Res.}\ }\textbf {\bibinfo {volume} {7}},\ \bibinfo {pages} {031001} (\bibinfo {year} {2025})},\ \Eprint {https://arxiv.org/abs/2502.09742} {arXiv:2502.09742 [nucl-ex]} \BibitemShut {NoStop}%
\bibitem [{\citenamefont {Li}\ \emph {et~al.}(2026)\citenamefont {Li}, \citenamefont {Shou},\ and\ \citenamefont {Wang}}]{Li:2511.07358}%
  \BibitemOpen
  \bibfield  {author} {\bibinfo {author} {\bibfnamefont {W.}~\bibnamefont {Li}}, \bibinfo {author} {\bibfnamefont {Q.}~\bibnamefont {Shou}},\ and\ \bibinfo {author} {\bibfnamefont {F.}~\bibnamefont {Wang}},\ }\href {https://doi.org/10.1140/epjs/s11734-026-02225-x} {\bibinfo {title} {Experimental review on the chiral magnetic effect in relativistic heavy ion collisions}},\ \bibinfo {howpublished} {Eur. Phys. J. Spec. Top.} (\bibinfo {year} {2026}),\ \Eprint {https://arxiv.org/abs/2511.07358} {2511.07358} \BibitemShut {NoStop}%
\bibitem [{\citenamefont {{STAR~Collaboration}}(2025)}]{STAR:2506.00275}%
  \BibitemOpen
  \bibfield  {author} {\bibinfo {author} {\bibnamefont {{STAR~Collaboration}}},\ }\href@noop {} {\bibinfo {title} {Charge separation measurements in {Au+Au} collisions at $\sqrt{s_{NN}}= 7.7-200$ gev in search of the {Chiral Magnetic Effect}}} (\bibinfo {year} {2025}),\ \Eprint {https://arxiv.org/abs/2506.00275} {2506.00275} \BibitemShut {NoStop}%
\bibitem [{\citenamefont {Ikeda}\ \emph {et~al.}(2021)\citenamefont {Ikeda}, \citenamefont {Kharzeev},\ and\ \citenamefont {Kikuchi}}]{Ikeda:2012.02926}%
  \BibitemOpen
  \bibfield  {author} {\bibinfo {author} {\bibfnamefont {K.}~\bibnamefont {Ikeda}}, \bibinfo {author} {\bibfnamefont {D.~E.}\ \bibnamefont {Kharzeev}},\ and\ \bibinfo {author} {\bibfnamefont {Y.}~\bibnamefont {Kikuchi}},\ }\href {https://doi.org/10.1103/PhysRevD.103.L071502} {\bibfield  {journal} {\bibinfo  {journal} {Phys. Rev. D}\ }\textbf {\bibinfo {volume} {103}},\ \bibinfo {pages} {L071502} (\bibinfo {year} {2021})},\ \Eprint {https://arxiv.org/abs/2012.02926} {arXiv:2012.02926 [hep-ph]} \BibitemShut {NoStop}%
\bibitem [{\citenamefont {Fej{\H{o}}s}\ and\ \citenamefont {Suenaga}(2025)}]{Fejos:2506.14010}%
  \BibitemOpen
  \bibfield  {author} {\bibinfo {author} {\bibfnamefont {G.}~\bibnamefont {Fej{\H{o}}s}}\ and\ \bibinfo {author} {\bibfnamefont {D.}~\bibnamefont {Suenaga}},\ }\href {https://doi.org/https://dx.doi.org/10.1103/s79l-gkt2} {\bibfield  {journal} {\bibinfo  {journal} {Phys.~Rev.~D}\ }\textbf {\bibinfo {volume} {112}},\ \bibinfo {pages} {114028} (\bibinfo {year} {2025})},\ \Eprint {https://arxiv.org/abs/2506.14010} {2506.14010} \BibitemShut {NoStop}%
\bibitem [{\citenamefont {Gynther}\ \emph {et~al.}(2011)\citenamefont {Gynther}, \citenamefont {Landsteiner}, \citenamefont {{Pena-Benitez}},\ and\ \citenamefont {Rebhan}}]{Gynther:1005.2587}%
  \BibitemOpen
  \bibfield  {author} {\bibinfo {author} {\bibfnamefont {A.}~\bibnamefont {Gynther}}, \bibinfo {author} {\bibfnamefont {K.}~\bibnamefont {Landsteiner}}, \bibinfo {author} {\bibfnamefont {F.}~\bibnamefont {{Pena-Benitez}}},\ and\ \bibinfo {author} {\bibfnamefont {A.}~\bibnamefont {Rebhan}},\ }\href {http://dx.doi.org/10.1007/JHEP02(2011)110} {\bibfield  {journal} {\bibinfo  {journal} {JHEP}\ }\textbf {\bibinfo {volume} {1102}},\ \bibinfo {pages} {110}},\ \Eprint {https://arxiv.org/abs/1005.2587} {1005.2587} \BibitemShut {NoStop}%
\bibitem [{\citenamefont {Rebhan}\ \emph {et~al.}(2010)\citenamefont {Rebhan}, \citenamefont {Schmitt},\ and\ \citenamefont {Stricker}}]{Rebhan:0909.4782}%
  \BibitemOpen
  \bibfield  {author} {\bibinfo {author} {\bibfnamefont {A.}~\bibnamefont {Rebhan}}, \bibinfo {author} {\bibfnamefont {A.}~\bibnamefont {Schmitt}},\ and\ \bibinfo {author} {\bibfnamefont {S.~A.}\ \bibnamefont {Stricker}},\ }\href {https://doi.org/10.1007/JHEP01(2010)026} {\bibfield  {journal} {\bibinfo  {journal} {JHEP}\ }\textbf {\bibinfo {volume} {1001}},\ \bibinfo {pages} {026}},\ \Eprint {https://arxiv.org/abs/0909.4782} {0909.4782} \BibitemShut {NoStop}%
\bibitem [{\citenamefont {Buividovich}\ \emph {et~al.}(2009)\citenamefont {Buividovich}, \citenamefont {Chernodub}, \citenamefont {Luschevskaya},\ and\ \citenamefont {Polikarpov}}]{Buividovich:09:7}%
  \BibitemOpen
  \bibfield  {author} {\bibinfo {author} {\bibfnamefont {P.~V.}\ \bibnamefont {Buividovich}}, \bibinfo {author} {\bibfnamefont {M.~N.}\ \bibnamefont {Chernodub}}, \bibinfo {author} {\bibfnamefont {E.~V.}\ \bibnamefont {Luschevskaya}},\ and\ \bibinfo {author} {\bibfnamefont {M.~I.}\ \bibnamefont {Polikarpov}},\ }\href {https://doi.org/10.1103/PhysRevD.80.054503} {\bibfield  {journal} {\bibinfo  {journal} {Phys.~Rev.~D}\ }\textbf {\bibinfo {volume} {80}},\ \bibinfo {pages} {054503} (\bibinfo {year} {2009})},\ \Eprint {https://arxiv.org/abs/0907.0494} {0907.0494} \BibitemShut {NoStop}%
\bibitem [{\citenamefont {Yamamoto}(2011{\natexlab{a}})}]{Yamamoto:1111.4681}%
  \BibitemOpen
  \bibfield  {author} {\bibinfo {author} {\bibfnamefont {A.}~\bibnamefont {Yamamoto}},\ }\href {https://doi.org/10.1103/PhysRevD.84.114504} {\bibfield  {journal} {\bibinfo  {journal} {Phys.~Rev.~D}\ }\textbf {\bibinfo {volume} {84}},\ \bibinfo {pages} {114504} (\bibinfo {year} {2011}{\natexlab{a}})},\ \Eprint {https://arxiv.org/abs/1111.4681} {1111.4681} \BibitemShut {NoStop}%
\bibitem [{\citenamefont {Yamamoto}(2011{\natexlab{b}})}]{Yamamoto:1105.0385}%
  \BibitemOpen
  \bibfield  {author} {\bibinfo {author} {\bibfnamefont {A.}~\bibnamefont {Yamamoto}},\ }\href {http://dx.doi.org/10.1103/PhysRevLett.107.031601} {\bibfield  {journal} {\bibinfo  {journal} {Phys.~Rev.~Lett.}\ }\textbf {\bibinfo {volume} {107}},\ \bibinfo {pages} {031601} (\bibinfo {year} {2011}{\natexlab{b}})},\ \Eprint {https://arxiv.org/abs/1105.0385} {1105.0385} \BibitemShut {NoStop}%
\bibitem [{\citenamefont {Brandt}\ \emph {et~al.}(2024)\citenamefont {Brandt}, \citenamefont {Endrődi}, \citenamefont {{Garnacho-Velasco}},\ and\ \citenamefont {Markó}}]{Brandt:2405.09484}%
  \BibitemOpen
  \bibfield  {author} {\bibinfo {author} {\bibfnamefont {B.~B.}\ \bibnamefont {Brandt}}, \bibinfo {author} {\bibfnamefont {G.}~\bibnamefont {Endrődi}}, \bibinfo {author} {\bibfnamefont {E.}~\bibnamefont {{Garnacho-Velasco}}},\ and\ \bibinfo {author} {\bibfnamefont {G.}~\bibnamefont {Markó}},\ }\href {https://doi.org/10.1007/JHEP09(2024)092} {\bibfield  {journal} {\bibinfo  {journal} {JHEP}\ }\textbf {\bibinfo {volume} {2409}},\ \bibinfo {pages} {092}},\ \Eprint {https://arxiv.org/abs/2405.09484} {2405.09484} \BibitemShut {NoStop}%
\bibitem [{\citenamefont {Buividovich}\ \emph {et~al.}(2020)\citenamefont {Buividovich}, \citenamefont {Smith},\ and\ \citenamefont {{von~Smekal}}}]{Buividovich:20:1}%
  \BibitemOpen
  \bibfield  {author} {\bibinfo {author} {\bibfnamefont {P.~V.}\ \bibnamefont {Buividovich}}, \bibinfo {author} {\bibfnamefont {D.}~\bibnamefont {Smith}},\ and\ \bibinfo {author} {\bibfnamefont {L.}~\bibnamefont {{von~Smekal}}},\ }\href {http://dx.doi.org/10.1103/PhysRevD.102.094510} {\bibfield  {journal} {\bibinfo  {journal} {Phys.~Rev.~D}\ }\textbf {\bibinfo {volume} {102}},\ \bibinfo {pages} {094510} (\bibinfo {year} {2020})},\ \Eprint {https://arxiv.org/abs/2007.05639} {2007.05639} \BibitemShut {NoStop}%
\bibitem [{\citenamefont {Braguta}\ \emph {et~al.}(2016)\citenamefont {Braguta}, \citenamefont {Ilgenfritz}, \citenamefont {Kotov}, \citenamefont {Molochkov},\ and\ \citenamefont {Nikolaev}}]{Braguta:1605.04090}%
  \BibitemOpen
  \bibfield  {author} {\bibinfo {author} {\bibfnamefont {V.~V.}\ \bibnamefont {Braguta}}, \bibinfo {author} {\bibfnamefont {E.}~\bibnamefont {Ilgenfritz}}, \bibinfo {author} {\bibfnamefont {A.~Y.}\ \bibnamefont {Kotov}}, \bibinfo {author} {\bibfnamefont {A.~V.}\ \bibnamefont {Molochkov}},\ and\ \bibinfo {author} {\bibfnamefont {A.~A.}\ \bibnamefont {Nikolaev}},\ }\href {http://dx.doi.org/10.1103/PhysRevD.94.114510} {\bibfield  {journal} {\bibinfo  {journal} {Phys.~Rev.~D}\ }\textbf {\bibinfo {volume} {94}},\ \bibinfo {pages} {114510} (\bibinfo {year} {2016})},\ \Eprint {https://arxiv.org/abs/1605.04090} {1605.04090} \BibitemShut {NoStop}%
\bibitem [{\citenamefont {Wilhelm}\ \emph {et~al.}(2019)\citenamefont {Wilhelm}, \citenamefont {Holicki}, \citenamefont {Smith}, \citenamefont {Wellegehausen},\ and\ \citenamefont {{von Smekal}}}]{Smith:1910.04495}%
  \BibitemOpen
  \bibfield  {author} {\bibinfo {author} {\bibfnamefont {J.}~\bibnamefont {Wilhelm}}, \bibinfo {author} {\bibfnamefont {L.}~\bibnamefont {Holicki}}, \bibinfo {author} {\bibfnamefont {D.}~\bibnamefont {Smith}}, \bibinfo {author} {\bibfnamefont {B.}~\bibnamefont {Wellegehausen}},\ and\ \bibinfo {author} {\bibfnamefont {L.}~\bibnamefont {{von Smekal}}},\ }\href {http://dx.doi.org/10.1103/PhysRevD.100.114507} {\bibfield  {journal} {\bibinfo  {journal} {Phys.~Rev.~D}\ }\textbf {\bibinfo {volume} {100}},\ \bibinfo {pages} {114507} (\bibinfo {year} {2019})},\ \Eprint {https://arxiv.org/abs/1910.04495} {1910.04495} \BibitemShut {NoStop}%
\bibitem [{\citenamefont {Lawlor}\ \emph {et~al.}(2022)\citenamefont {Lawlor}, \citenamefont {Hands}, \citenamefont {Kim},\ and\ \citenamefont {Skullerud}}]{Lawlor:2210.07731}%
  \BibitemOpen
  \bibfield  {author} {\bibinfo {author} {\bibfnamefont {D.}~\bibnamefont {Lawlor}}, \bibinfo {author} {\bibfnamefont {S.}~\bibnamefont {Hands}}, \bibinfo {author} {\bibfnamefont {S.}~\bibnamefont {Kim}},\ and\ \bibinfo {author} {\bibfnamefont {J.-I.}\ \bibnamefont {Skullerud}},\ }\href {https://doi.org/10.1051/epjconf/202227407012} {\bibfield  {journal} {\bibinfo  {journal} {EPJ Web Conf.}\ }\textbf {\bibinfo {volume} {274}},\ \bibinfo {pages} {07012} (\bibinfo {year} {2022})},\ \Eprint {https://arxiv.org/abs/2210.07731} {arXiv:2210.07731 [hep-lat]} \BibitemShut {NoStop}%
\bibitem [{\citenamefont {Kogut}\ \emph {et~al.}(2001{\natexlab{a}})\citenamefont {Kogut}, \citenamefont {Sinclair}, \citenamefont {Hands},\ and\ \citenamefont {Morrison}}]{Kogut:hep-lat/0105026}%
  \BibitemOpen
  \bibfield  {author} {\bibinfo {author} {\bibfnamefont {J.~B.}\ \bibnamefont {Kogut}}, \bibinfo {author} {\bibfnamefont {D.~K.}\ \bibnamefont {Sinclair}}, \bibinfo {author} {\bibfnamefont {S.~J.}\ \bibnamefont {Hands}},\ and\ \bibinfo {author} {\bibfnamefont {S.~E.}\ \bibnamefont {Morrison}},\ }\href {http://dx.doi.org/10.1103/PhysRevD.64.094505} {\bibfield  {journal} {\bibinfo  {journal} {Phys.~Rev.~D}\ }\textbf {\bibinfo {volume} {64}},\ \bibinfo {pages} {094505} (\bibinfo {year} {2001}{\natexlab{a}})},\ \Eprint {https://arxiv.org/abs/hep-lat/0105026} {hep-lat/0105026} \BibitemShut {NoStop}%
\bibitem [{\citenamefont {Kogut}\ \emph {et~al.}(2000)\citenamefont {Kogut}, \citenamefont {Stephanov}, \citenamefont {Toublan}, \citenamefont {Verbaarschot},\ and\ \citenamefont {Zhitnitsky}}]{Kogut:hep-ph/0001171}%
  \BibitemOpen
  \bibfield  {author} {\bibinfo {author} {\bibfnamefont {J.~B.}\ \bibnamefont {Kogut}}, \bibinfo {author} {\bibfnamefont {M.~A.}\ \bibnamefont {Stephanov}}, \bibinfo {author} {\bibfnamefont {D.}~\bibnamefont {Toublan}}, \bibinfo {author} {\bibfnamefont {J.~J.~M.}\ \bibnamefont {Verbaarschot}},\ and\ \bibinfo {author} {\bibfnamefont {A.}~\bibnamefont {Zhitnitsky}},\ }\href {http://dx.doi.org/10.1016/S0550-3213(00)00242-X} {\bibfield  {journal} {\bibinfo  {journal} {Nucl.~Phys.~B}\ }\textbf {\bibinfo {volume} {582}},\ \bibinfo {pages} {477 } (\bibinfo {year} {2000})},\ \Eprint {https://arxiv.org/abs/hep-ph/0001171} {hep-ph/0001171} \BibitemShut {NoStop}%
\bibitem [{\citenamefont {Iida}\ \emph {et~al.}(2026)\citenamefont {Iida}, \citenamefont {Itou}, \citenamefont {Murakami},\ and\ \citenamefont {Suenaga}}]{Suenaga:2606.13974}%
  \BibitemOpen
  \bibfield  {author} {\bibinfo {author} {\bibfnamefont {K.}~\bibnamefont {Iida}}, \bibinfo {author} {\bibfnamefont {E.}~\bibnamefont {Itou}}, \bibinfo {author} {\bibfnamefont {K.}~\bibnamefont {Murakami}},\ and\ \bibinfo {author} {\bibfnamefont {D.}~\bibnamefont {Suenaga}},\ }\href@noop {} {\bibinfo {title} {{Hadron spectra of finite-density QC$_2$D}}} (\bibinfo {year} {2026}),\ \Eprint {https://arxiv.org/abs/2606.13974} {2606.13974} \BibitemShut {NoStop}%
\bibitem [{\citenamefont {Hands}\ \emph {et~al.}(2006)\citenamefont {Hands}, \citenamefont {Kim},\ and\ \citenamefont {Skullerud}}]{Hands:hep-lat/0604004}%
  \BibitemOpen
  \bibfield  {author} {\bibinfo {author} {\bibfnamefont {S.}~\bibnamefont {Hands}}, \bibinfo {author} {\bibfnamefont {S.}~\bibnamefont {Kim}},\ and\ \bibinfo {author} {\bibfnamefont {J.}~\bibnamefont {Skullerud}},\ }\href {http://dx.doi.org/10.1140/epjc/s2006-02621-8} {\bibfield  {journal} {\bibinfo  {journal} {Eur.~Phys.~J.~C}\ }\textbf {\bibinfo {volume} {48}},\ \bibinfo {pages} {193} (\bibinfo {year} {2006})},\ \Eprint {https://arxiv.org/abs/hep-lat/0604004} {hep-lat/0604004} \BibitemShut {NoStop}%
\bibitem [{\citenamefont {Hands}\ \emph {et~al.}(2010)\citenamefont {Hands}, \citenamefont {Kim},\ and\ \citenamefont {Skullerud}}]{Hands:1001.1682}%
  \BibitemOpen
  \bibfield  {author} {\bibinfo {author} {\bibfnamefont {S.}~\bibnamefont {Hands}}, \bibinfo {author} {\bibfnamefont {S.}~\bibnamefont {Kim}},\ and\ \bibinfo {author} {\bibfnamefont {J.}~\bibnamefont {Skullerud}},\ }\href {http://dx.doi.org/10.1103/PhysRevD.81.091502} {\bibfield  {journal} {\bibinfo  {journal} {Phys.~Rev.~D}\ }\textbf {\bibinfo {volume} {81}},\ \bibinfo {pages} {091502(R)} (\bibinfo {year} {2010})},\ \Eprint {https://arxiv.org/abs/1001.1682} {1001.1682} \BibitemShut {NoStop}%
\bibitem [{\citenamefont {Strodthoff}\ \emph {et~al.}(2012)\citenamefont {Strodthoff}, \citenamefont {Schaefer},\ and\ \citenamefont {{von Smekal}}}]{Smekal:1112.5401}%
  \BibitemOpen
  \bibfield  {author} {\bibinfo {author} {\bibfnamefont {N.}~\bibnamefont {Strodthoff}}, \bibinfo {author} {\bibfnamefont {B.}~\bibnamefont {Schaefer}},\ and\ \bibinfo {author} {\bibfnamefont {L.}~\bibnamefont {{von Smekal}}},\ }\href {http://dx.doi.org/10.1103/PhysRevD.85.074007} {\bibfield  {journal} {\bibinfo  {journal} {Phys.~Rev.~D}\ }\textbf {\bibinfo {volume} {85}},\ \bibinfo {pages} {074007} (\bibinfo {year} {2012})},\ \Eprint {https://arxiv.org/abs/1112.5401} {1112.5401} \BibitemShut {NoStop}%
\bibitem [{\citenamefont {Cotter}\ \emph {et~al.}(2013)\citenamefont {Cotter}, \citenamefont {Giudice}, \citenamefont {Hands},\ and\ \citenamefont {Skullerud}}]{Hands:1210.4496}%
  \BibitemOpen
  \bibfield  {author} {\bibinfo {author} {\bibfnamefont {S.}~\bibnamefont {Cotter}}, \bibinfo {author} {\bibfnamefont {P.}~\bibnamefont {Giudice}}, \bibinfo {author} {\bibfnamefont {S.}~\bibnamefont {Hands}},\ and\ \bibinfo {author} {\bibfnamefont {J.}~\bibnamefont {Skullerud}},\ }\href {http://dx.doi.org/10.1103/PhysRevD.87.034507} {\bibfield  {journal} {\bibinfo  {journal} {Phys.~Rev.~D}\ }\textbf {\bibinfo {volume} {87}},\ \bibinfo {pages} {034507} (\bibinfo {year} {2013})},\ \Eprint {https://arxiv.org/abs/1210.4496} {1210.4496} \BibitemShut {NoStop}%
\bibitem [{\citenamefont {Strodthoff}\ and\ \citenamefont {{von Smekal}}(2014)}]{Strodthoff:1306.2897}%
  \BibitemOpen
  \bibfield  {author} {\bibinfo {author} {\bibfnamefont {N.}~\bibnamefont {Strodthoff}}\ and\ \bibinfo {author} {\bibfnamefont {L.}~\bibnamefont {{von Smekal}}},\ }\href {http://dx.doi.org/10.1016/j.physletb.2014.03.008} {\bibfield  {journal} {\bibinfo  {journal} {Phys.~Lett.~B}\ }\textbf {\bibinfo {volume} {731}},\ \bibinfo {pages} {350 } (\bibinfo {year} {2014})},\ \Eprint {https://arxiv.org/abs/1306.2897} {1306.2897} \BibitemShut {NoStop}%
\bibitem [{\citenamefont {Boz}\ \emph {et~al.}(2016)\citenamefont {Boz}, \citenamefont {Giudice}, \citenamefont {Hands}, \citenamefont {Skullerud},\ and\ \citenamefont {Williams}}]{Hands:1502.01219}%
  \BibitemOpen
  \bibfield  {author} {\bibinfo {author} {\bibfnamefont {T.}~\bibnamefont {Boz}}, \bibinfo {author} {\bibfnamefont {P.}~\bibnamefont {Giudice}}, \bibinfo {author} {\bibfnamefont {S.}~\bibnamefont {Hands}}, \bibinfo {author} {\bibfnamefont {J.}~\bibnamefont {Skullerud}},\ and\ \bibinfo {author} {\bibfnamefont {A.~G.}\ \bibnamefont {Williams}},\ }\href {http://dx.doi.org/10.1063/1.4938682} {\bibfield  {journal} {\bibinfo  {journal} {AIP~Conf.~Proc.}\ }\textbf {\bibinfo {volume} {1701}},\ \bibinfo {pages} {060019} (\bibinfo {year} {2016})},\ \Eprint {https://arxiv.org/abs/1502.01219} {1502.01219} \BibitemShut {NoStop}%
\bibitem [{\citenamefont {Bornyakov}\ \emph {et~al.}(2018)\citenamefont {Bornyakov}, \citenamefont {Braguta}, \citenamefont {Ilgenfritz}, \citenamefont {Kotov}, \citenamefont {Molochkov},\ and\ \citenamefont {Nikolaev}}]{Braguta:1711.01869}%
  \BibitemOpen
  \bibfield  {author} {\bibinfo {author} {\bibfnamefont {V.~G.}\ \bibnamefont {Bornyakov}}, \bibinfo {author} {\bibfnamefont {V.~V.}\ \bibnamefont {Braguta}}, \bibinfo {author} {\bibfnamefont {E.}~\bibnamefont {Ilgenfritz}}, \bibinfo {author} {\bibfnamefont {A.~Y.}\ \bibnamefont {Kotov}}, \bibinfo {author} {\bibfnamefont {A.~V.}\ \bibnamefont {Molochkov}},\ and\ \bibinfo {author} {\bibfnamefont {A.~A.}\ \bibnamefont {Nikolaev}},\ }\href {http://dx.doi.org/10.1007/JHEP03(2018)161} {\bibfield  {journal} {\bibinfo  {journal} {JHEP}\ }\textbf {\bibinfo {volume} {1803}},\ \bibinfo {pages} {161}},\ \Eprint {https://arxiv.org/abs/1711.01869} {1711.01869} \BibitemShut {NoStop}%
\bibitem [{\citenamefont {Holicki}\ \emph {et~al.}(2017)\citenamefont {Holicki}, \citenamefont {Wilhelm}, \citenamefont {Smith}, \citenamefont {Wellegehausen},\ and\ \citenamefont {{von Smekal}}}]{Holicki:1701.04664}%
  \BibitemOpen
  \bibfield  {author} {\bibinfo {author} {\bibfnamefont {L.}~\bibnamefont {Holicki}}, \bibinfo {author} {\bibfnamefont {J.}~\bibnamefont {Wilhelm}}, \bibinfo {author} {\bibfnamefont {D.}~\bibnamefont {Smith}}, \bibinfo {author} {\bibfnamefont {B.}~\bibnamefont {Wellegehausen}},\ and\ \bibinfo {author} {\bibfnamefont {L.}~\bibnamefont {{von Smekal}}},\ }\href {https://pos.sissa.it/256/052/} {\bibfield  {journal} {\bibinfo  {journal} {PoS}\ }\textbf {\bibinfo {volume} {LATTICE2016}},\ \bibinfo {pages} {052} (\bibinfo {year} {2017})},\ \Eprint {https://arxiv.org/abs/1701.04664} {1701.04664} \BibitemShut {NoStop}%
\bibitem [{\citenamefont {Iida}\ \emph {et~al.}(2020)\citenamefont {Iida}, \citenamefont {Itou},\ and\ \citenamefont {Lee}}]{Etou:1910.07872}%
  \BibitemOpen
  \bibfield  {author} {\bibinfo {author} {\bibfnamefont {K.}~\bibnamefont {Iida}}, \bibinfo {author} {\bibfnamefont {E.}~\bibnamefont {Itou}},\ and\ \bibinfo {author} {\bibfnamefont {T.}~\bibnamefont {Lee}},\ }\href {http://dx.doi.org/10.1007/JHEP01(2020)181} {\bibfield  {journal} {\bibinfo  {journal} {JHEP}\ }\textbf {\bibinfo {volume} {01}},\ \bibinfo {pages} {181}},\ \Eprint {https://arxiv.org/abs/1910.07872} {1910.07872} \BibitemShut {NoStop}%
\bibitem [{\citenamefont {Contant}\ and\ \citenamefont {Huber}(2020)}]{Huber:1909.12796}%
  \BibitemOpen
  \bibfield  {author} {\bibinfo {author} {\bibfnamefont {R.}~\bibnamefont {Contant}}\ and\ \bibinfo {author} {\bibfnamefont {M.~Q.}\ \bibnamefont {Huber}},\ }\href {http://dx.doi.org/10.1103/PhysRevD.101.014016} {\bibfield  {journal} {\bibinfo  {journal} {Phys.~Rev.~D}\ }\textbf {\bibinfo {volume} {101}},\ \bibinfo {pages} {014016} (\bibinfo {year} {2020})},\ \Eprint {https://arxiv.org/abs/1909.12796} {1909.12796} \BibitemShut {NoStop}%
\bibitem [{\citenamefont {Boz}\ \emph {et~al.}(2020)\citenamefont {Boz}, \citenamefont {Giudice}, \citenamefont {Hands},\ and\ \citenamefont {Skullerud}}]{Hands:1912.10975}%
  \BibitemOpen
  \bibfield  {author} {\bibinfo {author} {\bibfnamefont {T.}~\bibnamefont {Boz}}, \bibinfo {author} {\bibfnamefont {P.}~\bibnamefont {Giudice}}, \bibinfo {author} {\bibfnamefont {S.}~\bibnamefont {Hands}},\ and\ \bibinfo {author} {\bibfnamefont {J.}~\bibnamefont {Skullerud}},\ }\href {http://dx.doi.org/10.1103/PhysRevD.101.074506} {\bibfield  {journal} {\bibinfo  {journal} {Phys.~Rev.~D}\ }\textbf {\bibinfo {volume} {101}},\ \bibinfo {pages} {074506} (\bibinfo {year} {2020})},\ \Eprint {https://arxiv.org/abs/1912.10975} {1912.10975} \BibitemShut {NoStop}%
\bibitem [{\citenamefont {McLerran}\ and\ \citenamefont {Pisarski}(2007)}]{Pisarski:0706.2191}%
  \BibitemOpen
  \bibfield  {author} {\bibinfo {author} {\bibfnamefont {L.}~\bibnamefont {McLerran}}\ and\ \bibinfo {author} {\bibfnamefont {R.~D.}\ \bibnamefont {Pisarski}},\ }\href {http://dx.doi.org/10.1016/j.nuclphysa.2007.08.013} {\bibfield  {journal} {\bibinfo  {journal} {Nucl.~Phys.~A}\ }\textbf {\bibinfo {volume} {796}},\ \bibinfo {pages} {83 } (\bibinfo {year} {2007})},\ \Eprint {https://arxiv.org/abs/0706.2191} {0706.2191} \BibitemShut {NoStop}%
\bibitem [{\citenamefont {Endr\H{o}di}(2014)}]{Endrodi:1407.1216}%
  \BibitemOpen
  \bibfield  {author} {\bibinfo {author} {\bibfnamefont {G.}~\bibnamefont {Endr\H{o}di}},\ }\href {https://dx.doi.org/10.1103/PhysRevD.90.094501} {\bibfield  {journal} {\bibinfo  {journal} {Phys.~Rev.~D}\ }\textbf {\bibinfo {volume} {90}},\ \bibinfo {pages} {094501} (\bibinfo {year} {2014})},\ \Eprint {https://arxiv.org/abs/1407.1216} {1407.1216} \BibitemShut {NoStop}%
\bibitem [{\citenamefont {Brandt}\ \emph {et~al.}(2025{\natexlab{a}})\citenamefont {Brandt}, \citenamefont {Chelnokov}, \citenamefont {Endrodi}, \citenamefont {Marko}, \citenamefont {Scheid},\ and\ \citenamefont {{von~Smekal}}}]{Smekal:2502.04025}%
  \BibitemOpen
  \bibfield  {author} {\bibinfo {author} {\bibfnamefont {B.~B.}\ \bibnamefont {Brandt}}, \bibinfo {author} {\bibfnamefont {V.}~\bibnamefont {Chelnokov}}, \bibinfo {author} {\bibfnamefont {G.}~\bibnamefont {Endrodi}}, \bibinfo {author} {\bibfnamefont {G.}~\bibnamefont {Marko}}, \bibinfo {author} {\bibfnamefont {D.}~\bibnamefont {Scheid}},\ and\ \bibinfo {author} {\bibfnamefont {L.}~\bibnamefont {{von~Smekal}}},\ }\href@noop {} {\bibinfo {title} {Renormalization group invariant mean-field model for {QCD} at finite isospin density}} (\bibinfo {year} {2025}{\natexlab{a}}),\ \Eprint {https://arxiv.org/abs/2502.04025} {2502.04025} \BibitemShut {NoStop}%
\bibitem [{\citenamefont {Cohen}(2003)}]{Cohen:hep-ph/0304024}%
  \BibitemOpen
  \bibfield  {author} {\bibinfo {author} {\bibfnamefont {T.~D.}\ \bibnamefont {Cohen}},\ }\href {https://dx.doi.org/10.1103/PhysRevLett.91.032002} {\bibfield  {journal} {\bibinfo  {journal} {Phys.~Rev.~Lett.}\ }\textbf {\bibinfo {volume} {91}},\ \bibinfo {pages} {032002} (\bibinfo {year} {2003})},\ \Eprint {https://arxiv.org/abs/hep-ph/0304024} {hep-ph/0304024} \BibitemShut {NoStop}%
\bibitem [{\citenamefont {Fukushima}\ \emph {et~al.}(2010)\citenamefont {Fukushima}, \citenamefont {Kharzeev},\ and\ \citenamefont {Warringa}}]{Fukushima:0912.2961}%
  \BibitemOpen
  \bibfield  {author} {\bibinfo {author} {\bibfnamefont {K.}~\bibnamefont {Fukushima}}, \bibinfo {author} {\bibfnamefont {D.~E.}\ \bibnamefont {Kharzeev}},\ and\ \bibinfo {author} {\bibfnamefont {H.~J.}\ \bibnamefont {Warringa}},\ }\href {https://doi.org/10.1016/j.nuclphysa.2010.02.003} {\bibfield  {journal} {\bibinfo  {journal} {Nucl. Phys. A}\ }\textbf {\bibinfo {volume} {836}},\ \bibinfo {pages} {311} (\bibinfo {year} {2010})},\ \Eprint {https://arxiv.org/abs/0912.2961} {arXiv:0912.2961 [hep-ph]} \BibitemShut {NoStop}%
\bibitem [{\citenamefont {Buividovich}\ \emph {et~al.}(2010)\citenamefont {Buividovich}, \citenamefont {Chernodub}, \citenamefont {Kharzeev}, \citenamefont {Kalaydzhyan}, \citenamefont {Luschevskaya},\ and\ \citenamefont {Polikarpov}}]{Buividovich:10:1}%
  \BibitemOpen
  \bibfield  {author} {\bibinfo {author} {\bibfnamefont {P.~V.}\ \bibnamefont {Buividovich}}, \bibinfo {author} {\bibfnamefont {M.~N.}\ \bibnamefont {Chernodub}}, \bibinfo {author} {\bibfnamefont {D.~E.}\ \bibnamefont {Kharzeev}}, \bibinfo {author} {\bibfnamefont {T.}~\bibnamefont {Kalaydzhyan}}, \bibinfo {author} {\bibfnamefont {E.~V.}\ \bibnamefont {Luschevskaya}},\ and\ \bibinfo {author} {\bibfnamefont {M.~I.}\ \bibnamefont {Polikarpov}},\ }\href {https://dx.doi.org/10.1103/PhysRevLett.105.132001} {\bibfield  {journal} {\bibinfo  {journal} {Phys.~Rev.~Lett.}\ }\textbf {\bibinfo {volume} {105}},\ \bibinfo {pages} {132001} (\bibinfo {year} {2010})},\ \Eprint {https://arxiv.org/abs/1003.2180} {1003.2180} \BibitemShut {NoStop}%
\bibitem [{\citenamefont {Almirante}\ \emph {et~al.}(2025)\citenamefont {Almirante}, \citenamefont {Astrakhantsev}, \citenamefont {Braguta}, \citenamefont {{D'Elia}}, \citenamefont {Maio}, \citenamefont {Naviglio}, \citenamefont {Sanfilippo},\ and\ \citenamefont {Trunin}}]{Braguta:2406.18504}%
  \BibitemOpen
  \bibfield  {author} {\bibinfo {author} {\bibfnamefont {G.}~\bibnamefont {Almirante}}, \bibinfo {author} {\bibfnamefont {N.}~\bibnamefont {Astrakhantsev}}, \bibinfo {author} {\bibfnamefont {V.~V.}\ \bibnamefont {Braguta}}, \bibinfo {author} {\bibfnamefont {M.}~\bibnamefont {{D'Elia}}}, \bibinfo {author} {\bibfnamefont {L.}~\bibnamefont {Maio}}, \bibinfo {author} {\bibfnamefont {M.}~\bibnamefont {Naviglio}}, \bibinfo {author} {\bibfnamefont {F.}~\bibnamefont {Sanfilippo}},\ and\ \bibinfo {author} {\bibfnamefont {A.}~\bibnamefont {Trunin}},\ }\href {https://doi.org/10.1103/PhysRevD.111.034505} {\bibfield  {journal} {\bibinfo  {journal} {Phys.~Rev.~D}\ }\textbf {\bibinfo {volume} {111}},\ \bibinfo {pages} {034505} (\bibinfo {year} {2025})},\ \Eprint {https://arxiv.org/abs/2406.18504} {2406.18504} \BibitemShut {NoStop}%
\bibitem [{\citenamefont {Astrakhantsev}\ \emph {et~al.}(2020)\citenamefont {Astrakhantsev}, \citenamefont {Braguta}, \citenamefont {{D'Elia}}, \citenamefont {Kotov}, \citenamefont {Nikolaev},\ and\ \citenamefont {Sanfilippo}}]{Braguta:1910.08516}%
  \BibitemOpen
  \bibfield  {author} {\bibinfo {author} {\bibfnamefont {N.~Y.}\ \bibnamefont {Astrakhantsev}}, \bibinfo {author} {\bibfnamefont {V.~V.}\ \bibnamefont {Braguta}}, \bibinfo {author} {\bibfnamefont {M.}~\bibnamefont {{D'Elia}}}, \bibinfo {author} {\bibfnamefont {A.~Y.}\ \bibnamefont {Kotov}}, \bibinfo {author} {\bibfnamefont {A.~A.}\ \bibnamefont {Nikolaev}},\ and\ \bibinfo {author} {\bibfnamefont {F.}~\bibnamefont {Sanfilippo}},\ }\href {https://doi.org/10.1103/PhysRevD.102.054516} {\bibfield  {journal} {\bibinfo  {journal} {Phys.~Rev.~D}\ }\textbf {\bibinfo {volume} {102}},\ \bibinfo {pages} {054516} (\bibinfo {year} {2020})},\ \Eprint {https://arxiv.org/abs/1910.08516} {1910.08516} \BibitemShut {NoStop}%
\bibitem [{\citenamefont {Li}\ \emph {et~al.}(2016)\citenamefont {Li}, \citenamefont {Kharzeev}, \citenamefont {Zhang}, \citenamefont {Huang}, \citenamefont {Pletikosic}, \citenamefont {Fedorov}, \citenamefont {Zhong}, \citenamefont {Schneeloch}, \citenamefont {Gu},\ and\ \citenamefont {Valla}}]{Kharzeev:1412.6543}%
  \BibitemOpen
  \bibfield  {author} {\bibinfo {author} {\bibfnamefont {Q.}~\bibnamefont {Li}}, \bibinfo {author} {\bibfnamefont {D.~E.}\ \bibnamefont {Kharzeev}}, \bibinfo {author} {\bibfnamefont {C.}~\bibnamefont {Zhang}}, \bibinfo {author} {\bibfnamefont {Y.}~\bibnamefont {Huang}}, \bibinfo {author} {\bibfnamefont {I.}~\bibnamefont {Pletikosic}}, \bibinfo {author} {\bibfnamefont {A.~V.}\ \bibnamefont {Fedorov}}, \bibinfo {author} {\bibfnamefont {R.~D.}\ \bibnamefont {Zhong}}, \bibinfo {author} {\bibfnamefont {J.~A.}\ \bibnamefont {Schneeloch}}, \bibinfo {author} {\bibfnamefont {G.~D.}\ \bibnamefont {Gu}},\ and\ \bibinfo {author} {\bibfnamefont {T.}~\bibnamefont {Valla}},\ }\href {https://doi.org/10.1038/nphys3648} {\bibfield  {journal} {\bibinfo  {journal} {Nature Phys.}\ }\textbf {\bibinfo {volume} {12}},\ \bibinfo {pages} {550 } (\bibinfo {year} {2016})},\ \Eprint {https://arxiv.org/abs/1412.6543} {1412.6543} \BibitemShut {NoStop}%
\bibitem [{\citenamefont {Sun}\ and\ \citenamefont {Yang}(2016)}]{Sun:1603.02624}%
  \BibitemOpen
  \bibfield  {author} {\bibinfo {author} {\bibfnamefont {Y.-W.}\ \bibnamefont {Sun}}\ and\ \bibinfo {author} {\bibfnamefont {Q.}~\bibnamefont {Yang}},\ }\href {https://doi.org/10.1007/JHEP09(2016)122} {\bibfield  {journal} {\bibinfo  {journal} {JHEP}\ }\textbf {\bibinfo {volume} {09}},\ \bibinfo {pages} {122}},\ \Eprint {https://arxiv.org/abs/1603.02624} {arXiv:1603.02624 [hep-th]} \BibitemShut {NoStop}%
\bibitem [{\citenamefont {Fukushima}\ and\ \citenamefont {Okutsu}(2022)}]{Fukushima:2106.07968}%
  \BibitemOpen
  \bibfield  {author} {\bibinfo {author} {\bibfnamefont {K.}~\bibnamefont {Fukushima}}\ and\ \bibinfo {author} {\bibfnamefont {A.}~\bibnamefont {Okutsu}},\ }\href {https://doi.org/10.1103/PhysRevD.105.054016} {\bibfield  {journal} {\bibinfo  {journal} {Phys. Rev. D}\ }\textbf {\bibinfo {volume} {105}},\ \bibinfo {pages} {054016} (\bibinfo {year} {2022})},\ \Eprint {https://arxiv.org/abs/2106.07968} {arXiv:2106.07968 [hep-ph]} \BibitemShut {NoStop}%
\bibitem [{\citenamefont {Boyda}\ \emph {et~al.}(2017)\citenamefont {Boyda}, \citenamefont {Braguta}, \citenamefont {Katsnelson},\ and\ \citenamefont {Kotov}}]{Braguta:1707.09810}%
  \BibitemOpen
  \bibfield  {author} {\bibinfo {author} {\bibfnamefont {D.~L.}\ \bibnamefont {Boyda}}, \bibinfo {author} {\bibfnamefont {V.~V.}\ \bibnamefont {Braguta}}, \bibinfo {author} {\bibfnamefont {M.~I.}\ \bibnamefont {Katsnelson}},\ and\ \bibinfo {author} {\bibfnamefont {A.~Y.}\ \bibnamefont {Kotov}},\ }\href {https://doi.org/10.1016/j.aop.2018.07.006} {\bibfield  {journal} {\bibinfo  {journal} {Ann.~Phys.}\ }\textbf {\bibinfo {volume} {396}},\ \bibinfo {pages} {78 } (\bibinfo {year} {2017})},\ \Eprint {https://arxiv.org/abs/1707.09810} {1707.09810} \BibitemShut {NoStop}%
\bibitem [{\citenamefont {Buividovich}(2024)}]{Buividovich:24:2}%
  \BibitemOpen
  \bibfield  {author} {\bibinfo {author} {\bibfnamefont {P.~V.}\ \bibnamefont {Buividovich}},\ }\href@noop {} {\bibfield  {journal} {\bibinfo  {journal} {Phys.~Rev.~D}\ }\textbf {\bibinfo {volume} {110}},\ \bibinfo {pages} {094508} (\bibinfo {year} {2024})},\ \Eprint {https://arxiv.org/abs/2404.14263} {2404.14263} \BibitemShut {NoStop}%
\bibitem [{\citenamefont {Brandt}\ \emph {et~al.}(2025{\natexlab{b}})\citenamefont {Brandt}, \citenamefont {Endrodi}, \citenamefont {Garnacho~Velasco}, \citenamefont {Marko},\ and\ \citenamefont {Valois}}]{Brandt:2502.01155}%
  \BibitemOpen
  \bibfield  {author} {\bibinfo {author} {\bibfnamefont {B.~B.}\ \bibnamefont {Brandt}}, \bibinfo {author} {\bibfnamefont {G.}~\bibnamefont {Endrodi}}, \bibinfo {author} {\bibfnamefont {E.}~\bibnamefont {Garnacho~Velasco}}, \bibinfo {author} {\bibfnamefont {G.}~\bibnamefont {Marko}},\ and\ \bibinfo {author} {\bibfnamefont {A.~D.~M.}\ \bibnamefont {Valois}},\ }\href {https://doi.org/10.22323/1.466.0196} {\bibfield  {journal} {\bibinfo  {journal} {PoS}\ }\textbf {\bibinfo {volume} {LATTICE2024}},\ \bibinfo {pages} {196} (\bibinfo {year} {2025}{\natexlab{b}})},\ \Eprint {https://arxiv.org/abs/2502.01155} {arXiv:2502.01155 [hep-lat]} \BibitemShut {NoStop}%
\bibitem [{\citenamefont {Inghirami}\ \emph {et~al.}(2020)\citenamefont {Inghirami}, \citenamefont {Mace}, \citenamefont {Hirono}, \citenamefont {{Del Zanna}}, \citenamefont {Kharzeev},\ and\ \citenamefont {Bleicher}}]{Kharzeev:1908.07605}%
  \BibitemOpen
  \bibfield  {author} {\bibinfo {author} {\bibfnamefont {G.}~\bibnamefont {Inghirami}}, \bibinfo {author} {\bibfnamefont {M.}~\bibnamefont {Mace}}, \bibinfo {author} {\bibfnamefont {Y.}~\bibnamefont {Hirono}}, \bibinfo {author} {\bibfnamefont {L.}~\bibnamefont {{Del Zanna}}}, \bibinfo {author} {\bibfnamefont {D.~E.}\ \bibnamefont {Kharzeev}},\ and\ \bibinfo {author} {\bibfnamefont {M.}~\bibnamefont {Bleicher}},\ }\href {http://dx.doi.org/} {\bibfield  {journal} {\bibinfo  {journal} {Eur.~Phys.~J.~C}\ }\textbf {\bibinfo {volume} {80}},\ \bibinfo {pages} {293} (\bibinfo {year} {2020})},\ \Eprint {https://arxiv.org/abs/1908.07605} {1908.07605} \BibitemShut {NoStop}%
\bibitem [{\citenamefont {Voloshin}(2004)}]{Voloshin:hep-ph/0406311}%
  \BibitemOpen
  \bibfield  {author} {\bibinfo {author} {\bibfnamefont {S.~A.}\ \bibnamefont {Voloshin}},\ }\href {https://doi.org/10.1103/PhysRevC.70.057901} {\bibfield  {journal} {\bibinfo  {journal} {Phys.~Rev.~C}\ }\textbf {\bibinfo {volume} {70}},\ \bibinfo {pages} {057901} (\bibinfo {year} {2004})},\ \Eprint {https://arxiv.org/abs/hep-ph/0406311} {hep-ph/0406311} \BibitemShut {NoStop}%
\bibitem [{\citenamefont {Magdy}\ \emph {et~al.}(2018)\citenamefont {Magdy}, \citenamefont {Shi}, \citenamefont {Liao}, \citenamefont {Ajitanand},\ and\ \citenamefont {Lacey}}]{Magdy:1710.01717}%
  \BibitemOpen
  \bibfield  {author} {\bibinfo {author} {\bibfnamefont {N.}~\bibnamefont {Magdy}}, \bibinfo {author} {\bibfnamefont {S.}~\bibnamefont {Shi}}, \bibinfo {author} {\bibfnamefont {J.}~\bibnamefont {Liao}}, \bibinfo {author} {\bibfnamefont {N.}~\bibnamefont {Ajitanand}},\ and\ \bibinfo {author} {\bibfnamefont {R.~A.}\ \bibnamefont {Lacey}},\ }\href {https://doi.org/10.1103/PhysRevC.97.061901} {\bibfield  {journal} {\bibinfo  {journal} {Phys. Rev. C}\ }\textbf {\bibinfo {volume} {97}},\ \bibinfo {pages} {061901} (\bibinfo {year} {2018})},\ \Eprint {https://arxiv.org/abs/1710.01717} {arXiv:1710.01717 [physics.data-an]} \BibitemShut {NoStop}%
\bibitem [{\citenamefont {Yin}\ and\ \citenamefont {Liao}(2016)}]{Yin:1504.06906}%
  \BibitemOpen
  \bibfield  {author} {\bibinfo {author} {\bibfnamefont {Y.}~\bibnamefont {Yin}}\ and\ \bibinfo {author} {\bibfnamefont {J.}~\bibnamefont {Liao}},\ }\href {https://doi.org/10.1016/j.physletb.2016.02.065} {\bibfield  {journal} {\bibinfo  {journal} {Phys. Lett. B}\ }\textbf {\bibinfo {volume} {756}},\ \bibinfo {pages} {42} (\bibinfo {year} {2016})},\ \Eprint {https://arxiv.org/abs/1504.06906} {arXiv:1504.06906 [nucl-th]} \BibitemShut {NoStop}%
\bibitem [{\citenamefont {Zhang}\ and\ \citenamefont {Zubkov}(2019)}]{zhang2019hall}%
  \BibitemOpen
  \bibfield  {author} {\bibinfo {author} {\bibfnamefont {C.}~\bibnamefont {Zhang}}\ and\ \bibinfo {author} {\bibfnamefont {M.}~\bibnamefont {Zubkov}},\ }\href@noop {} {\bibfield  {journal} {\bibinfo  {journal} {JETP letters}\ }\textbf {\bibinfo {volume} {110}},\ \bibinfo {pages} {487} (\bibinfo {year} {2019})},\ \Eprint {https://arxiv.org/abs/1908.04138} {1908.04138} \BibitemShut {NoStop}%
\bibitem [{\citenamefont {Zhang}\ and\ \citenamefont {Zubkov}(2020)}]{zhang2020influence}%
  \BibitemOpen
  \bibfield  {author} {\bibinfo {author} {\bibfnamefont {C.}~\bibnamefont {Zhang}}\ and\ \bibinfo {author} {\bibfnamefont {M.}~\bibnamefont {Zubkov}},\ }\href@noop {} {\bibfield  {journal} {\bibinfo  {journal} {Journal of Physics A: Mathematical and Theoretical}\ }\textbf {\bibinfo {volume} {53}},\ \bibinfo {pages} {195002} (\bibinfo {year} {2020})},\ \Eprint {https://arxiv.org/abs/1902.06545} {1902.06545} \BibitemShut {NoStop}%
\bibitem [{\citenamefont {Zubkov}\ and\ \citenamefont {Abramchuk}(2023)}]{zubkov2023effect}%
  \BibitemOpen
  \bibfield  {author} {\bibinfo {author} {\bibfnamefont {M.~A.}\ \bibnamefont {Zubkov}}\ and\ \bibinfo {author} {\bibfnamefont {R.~A.}\ \bibnamefont {Abramchuk}},\ }\href {https://doi.org/10.1103/PhysRevD.107.094021} {\bibfield  {journal} {\bibinfo  {journal} {Physical Review D}\ }\textbf {\bibinfo {volume} {107}},\ \bibinfo {pages} {094021} (\bibinfo {year} {2023})},\ \Eprint {https://arxiv.org/abs/2301.12261} {2301.12261} \BibitemShut {NoStop}%
\bibitem [{\citenamefont {Buividovich}\ \emph {et~al.}(2021{\natexlab{a}})\citenamefont {Buividovich}, \citenamefont {Smith},\ and\ \citenamefont {{von~Smekal}}}]{Buividovich:20:2}%
  \BibitemOpen
  \bibfield  {author} {\bibinfo {author} {\bibfnamefont {P.~V.}\ \bibnamefont {Buividovich}}, \bibinfo {author} {\bibfnamefont {D.}~\bibnamefont {Smith}},\ and\ \bibinfo {author} {\bibfnamefont {L.}~\bibnamefont {{von~Smekal}}},\ }\href {http://dx.doi.org/10.1103/PhysRevD.104.014511} {\bibfield  {journal} {\bibinfo  {journal} {Phys.~Rev.~D}\ }\textbf {\bibinfo {volume} {104}},\ \bibinfo {pages} {014511} (\bibinfo {year} {2021}{\natexlab{a}})},\ \Eprint {https://arxiv.org/abs/2012.05184} {2012.05184} \BibitemShut {NoStop}%
\bibitem [{\citenamefont {Buividovich}\ \emph {et~al.}(2021{\natexlab{b}})\citenamefont {Buividovich}, \citenamefont {Smith},\ and\ \citenamefont {{von Smekal}}}]{Buividovich:21:1}%
  \BibitemOpen
  \bibfield  {author} {\bibinfo {author} {\bibfnamefont {P.~V.}\ \bibnamefont {Buividovich}}, \bibinfo {author} {\bibfnamefont {D.}~\bibnamefont {Smith}},\ and\ \bibinfo {author} {\bibfnamefont {L.}~\bibnamefont {{von Smekal}}},\ }\href {http://dx.doi.org/10.1140/epja/s10050-021-00604-7} {\bibfield  {journal} {\bibinfo  {journal} {Eur.~Phys.~J.~A}\ }\textbf {\bibinfo {volume} {57}},\ \bibinfo {pages} {293} (\bibinfo {year} {2021}{\natexlab{b}})},\ \Eprint {https://arxiv.org/abs/2104.10012} {2104.10012} \BibitemShut {NoStop}%
\bibitem [{\citenamefont {Scheffler}\ \emph {et~al.}(2013)\citenamefont {Scheffler}, \citenamefont {Schmidt}, \citenamefont {Smith},\ and\ \citenamefont {{von Smekal}}}]{Scheffler:1311.4324}%
  \BibitemOpen
  \bibfield  {author} {\bibinfo {author} {\bibfnamefont {D.}~\bibnamefont {Scheffler}}, \bibinfo {author} {\bibfnamefont {C.}~\bibnamefont {Schmidt}}, \bibinfo {author} {\bibfnamefont {D.}~\bibnamefont {Smith}},\ and\ \bibinfo {author} {\bibfnamefont {L.}~\bibnamefont {{von Smekal}}},\ }\href {https://pos.sissa.it/187/191/pdf} {\bibfield  {journal} {\bibinfo  {journal} {PoS}\ }\textbf {\bibinfo {volume} {LATTICE2013}},\ \bibinfo {pages} {191} (\bibinfo {year} {2013})},\ \Eprint {https://arxiv.org/abs/1311.4324} {1311.4324} \BibitemShut {NoStop}%
\bibitem [{\citenamefont {Kogut}\ \emph {et~al.}(2001{\natexlab{b}})\citenamefont {Kogut}, \citenamefont {Toublan},\ and\ \citenamefont {Sinclair}}]{Kogut:hep-lat/0104010}%
  \BibitemOpen
  \bibfield  {author} {\bibinfo {author} {\bibfnamefont {J.~B.}\ \bibnamefont {Kogut}}, \bibinfo {author} {\bibfnamefont {D.}~\bibnamefont {Toublan}},\ and\ \bibinfo {author} {\bibfnamefont {D.~K.}\ \bibnamefont {Sinclair}},\ }\href {http://dx.doi.org/10.1016/S0370-2693(01)00586-X} {\bibfield  {journal} {\bibinfo  {journal} {Phys.~Lett.~B}\ }\textbf {\bibinfo {volume} {514}},\ \bibinfo {pages} {77 } (\bibinfo {year} {2001}{\natexlab{b}})},\ \Eprint {https://arxiv.org/abs/hep-lat/0104010} {hep-lat/0104010} \BibitemShut {NoStop}%
\bibitem [{\citenamefont {Renner}\ \emph {et~al.}(2005)\citenamefont {Renner}, \citenamefont {Schroers}, \citenamefont {Edwards}, \citenamefont {Fleming}, \citenamefont {Hagler}, \citenamefont {Negele}, \citenamefont {Orginos}, \citenamefont {Pochinski},\ and\ \citenamefont {Richards}}]{Renner:hep-lat/0409130}%
  \BibitemOpen
  \bibfield  {author} {\bibinfo {author} {\bibfnamefont {D.~B.}\ \bibnamefont {Renner}}, \bibinfo {author} {\bibfnamefont {W.}~\bibnamefont {Schroers}}, \bibinfo {author} {\bibfnamefont {R.}~\bibnamefont {Edwards}}, \bibinfo {author} {\bibfnamefont {G.~T.}\ \bibnamefont {Fleming}}, \bibinfo {author} {\bibfnamefont {P.}~\bibnamefont {Hagler}}, \bibinfo {author} {\bibfnamefont {J.~W.}\ \bibnamefont {Negele}}, \bibinfo {author} {\bibfnamefont {K.}~\bibnamefont {Orginos}}, \bibinfo {author} {\bibfnamefont {A.~V.}\ \bibnamefont {Pochinski}},\ and\ \bibinfo {author} {\bibfnamefont {D.}~\bibnamefont {Richards}},\ }\href {http://dx.doi.org/10.1016/j.nuclphysbps.2004.11.357} {\bibfield  {journal} {\bibinfo  {journal} {Nucl.~Phys.~Proc.~Suppl.}\ }\textbf {\bibinfo {volume} {140}},\ \bibinfo {pages} {255 } (\bibinfo {year} {2005})},\ \Eprint {https://arxiv.org/abs/hep-lat/0409130} {hep-lat/0409130} \BibitemShut {NoStop}%
\bibitem [{\citenamefont {Edwards}\ \emph {et~al.}(2006)\citenamefont {Edwards}, \citenamefont {Fleming}, \citenamefont {Hagler}, \citenamefont {Negele}, \citenamefont {Orginos}, \citenamefont {Pochinsky}, \citenamefont {Renner}, \citenamefont {Richards},\ and\ \citenamefont {Schroers}}]{Edwards:hep-lat/0510062}%
  \BibitemOpen
  \bibfield  {author} {\bibinfo {author} {\bibfnamefont {R.~G.}\ \bibnamefont {Edwards}}, \bibinfo {author} {\bibfnamefont {G.~T.}\ \bibnamefont {Fleming}}, \bibinfo {author} {\bibfnamefont {P.}~\bibnamefont {Hagler}}, \bibinfo {author} {\bibfnamefont {J.~W.}\ \bibnamefont {Negele}}, \bibinfo {author} {\bibfnamefont {K.}~\bibnamefont {Orginos}}, \bibinfo {author} {\bibfnamefont {A.}~\bibnamefont {Pochinsky}}, \bibinfo {author} {\bibfnamefont {D.~B.}\ \bibnamefont {Renner}}, \bibinfo {author} {\bibfnamefont {D.~G.}\ \bibnamefont {Richards}},\ and\ \bibinfo {author} {\bibfnamefont {W.}~\bibnamefont {Schroers}},\ }\href {http://dx.doi.org/10.1103/PhysRevLett.96.052001} {\bibfield  {journal} {\bibinfo  {journal} {Phys.~Rev.~Lett.}\ }\textbf {\bibinfo {volume} {96}},\ \bibinfo {pages} {052001} (\bibinfo {year} {2006})},\ \Eprint {https://arxiv.org/abs/hep-lat/0510062} {hep-lat/0510062} \BibitemShut {NoStop}%
\bibitem [{\citenamefont {Berkowitz}\ \emph {et~al.}(2017)\citenamefont {Berkowitz}, \citenamefont {Brantley}, \citenamefont {Bouchard}, \citenamefont {Chang}, \citenamefont {Clark}, \citenamefont {Garron}, \citenamefont {Joo}, \citenamefont {Kurth}, \citenamefont {Monahan}, \citenamefont {{Monge-Camacho}}, \citenamefont {Nicholson}, \citenamefont {Orginos}, \citenamefont {Rinaldi}, \citenamefont {Vranas},\ and\ \citenamefont {{Walker-Loud}}}]{Berkowitz:1704.01114}%
  \BibitemOpen
  \bibfield  {author} {\bibinfo {author} {\bibfnamefont {E.}~\bibnamefont {Berkowitz}}, \bibinfo {author} {\bibfnamefont {D.}~\bibnamefont {Brantley}}, \bibinfo {author} {\bibfnamefont {C.}~\bibnamefont {Bouchard}}, \bibinfo {author} {\bibfnamefont {C.}~\bibnamefont {Chang}}, \bibinfo {author} {\bibfnamefont {M.~A.}\ \bibnamefont {Clark}}, \bibinfo {author} {\bibfnamefont {N.}~\bibnamefont {Garron}}, \bibinfo {author} {\bibfnamefont {B.}~\bibnamefont {Joo}}, \bibinfo {author} {\bibfnamefont {T.}~\bibnamefont {Kurth}}, \bibinfo {author} {\bibfnamefont {C.}~\bibnamefont {Monahan}}, \bibinfo {author} {\bibfnamefont {H.}~\bibnamefont {{Monge-Camacho}}}, \bibinfo {author} {\bibfnamefont {A.}~\bibnamefont {Nicholson}}, \bibinfo {author} {\bibfnamefont {K.}~\bibnamefont {Orginos}}, \bibinfo {author} {\bibfnamefont {E.}~\bibnamefont {Rinaldi}}, \bibinfo {author} {\bibfnamefont {P.}~\bibnamefont {Vranas}},\ and\ \bibinfo {author} {\bibfnamefont {A.}~\bibnamefont {{Walker-Loud}}},\ }\href@noop {} {\bibinfo {title} {An
  accurate calculation of the nucleon axial charge with lattice {QCD}}} (\bibinfo {year} {2017}),\ \Eprint {https://arxiv.org/abs/1704.01114} {1704.01114} \BibitemShut {NoStop}%
\bibitem [{\citenamefont {Horsley}\ \emph {et~al.}(2016)\citenamefont {Horsley}, \citenamefont {Nakamura}, \citenamefont {Perlt}, \citenamefont {Rakow}, \citenamefont {Schierholz}, \citenamefont {Schiller},\ and\ \citenamefont {Zanotti}}]{Rakow:1511.05304}%
  \BibitemOpen
  \bibfield  {author} {\bibinfo {author} {\bibfnamefont {R.}~\bibnamefont {Horsley}}, \bibinfo {author} {\bibfnamefont {Y.}~\bibnamefont {Nakamura}}, \bibinfo {author} {\bibfnamefont {H.}~\bibnamefont {Perlt}}, \bibinfo {author} {\bibfnamefont {P.~E.~L.}\ \bibnamefont {Rakow}}, \bibinfo {author} {\bibfnamefont {G.}~\bibnamefont {Schierholz}}, \bibinfo {author} {\bibfnamefont {A.}~\bibnamefont {Schiller}},\ and\ \bibinfo {author} {\bibfnamefont {J.~M.}\ \bibnamefont {Zanotti}},\ }\href {https://doi.org/10.22323/1.251.0138} {\bibfield  {journal} {\bibinfo  {journal} {PoS}\ }\textbf {\bibinfo {volume} {LAT2015}},\ \bibinfo {pages} {138} (\bibinfo {year} {2016})},\ \Eprint {https://arxiv.org/abs/1511.05304} {1511.05304} \BibitemShut {NoStop}%
\bibitem [{\citenamefont {Horsley}\ \emph {et~al.}(2017)\citenamefont {Horsley}, \citenamefont {Kazmin}, \citenamefont {Nakamura}, \citenamefont {Perlt}, \citenamefont {Rakow}, \citenamefont {Schierholz}, \citenamefont {Schiller},\ and\ \citenamefont {Zanotti}}]{Rakow:1612.04992}%
  \BibitemOpen
  \bibfield  {author} {\bibinfo {author} {\bibfnamefont {R.}~\bibnamefont {Horsley}}, \bibinfo {author} {\bibfnamefont {S.}~\bibnamefont {Kazmin}}, \bibinfo {author} {\bibfnamefont {Y.}~\bibnamefont {Nakamura}}, \bibinfo {author} {\bibfnamefont {H.}~\bibnamefont {Perlt}}, \bibinfo {author} {\bibfnamefont {P.~E.~L.}\ \bibnamefont {Rakow}}, \bibinfo {author} {\bibfnamefont {G.}~\bibnamefont {Schierholz}}, \bibinfo {author} {\bibfnamefont {A.}~\bibnamefont {Schiller}},\ and\ \bibinfo {author} {\bibfnamefont {J.~M.}\ \bibnamefont {Zanotti}},\ }\href {https://doi.org/10.22323/1.256.0149} {\bibfield  {journal} {\bibinfo  {journal} {PoS}\ }\textbf {\bibinfo {volume} {LAT2016}},\ \bibinfo {pages} {149} (\bibinfo {year} {2017})},\ \Eprint {https://arxiv.org/abs/1612.04992} {1612.04992} \BibitemShut {NoStop}%
\bibitem [{\citenamefont {Buividovich}(2013)}]{Buividovich:13:7}%
  \BibitemOpen
  \bibfield  {author} {\bibinfo {author} {\bibfnamefont {P.~V.}\ \bibnamefont {Buividovich}},\ }\href {https://dx.doi.org/10.1088/1742-6596/607/1/012018} {\bibfield  {journal} {\bibinfo  {journal} {J.~Phys.:~Conf.~Ser.}\ }\textbf {\bibinfo {volume} {607}},\ \bibinfo {pages} {012018} (\bibinfo {year} {2013})},\ \Eprint {https://arxiv.org/abs/1309.4966} {1309.4966} \BibitemShut {NoStop}%
\bibitem [{\citenamefont {Buividovich}(2014)}]{Buividovich:13:8}%
  \BibitemOpen
  \bibfield  {author} {\bibinfo {author} {\bibfnamefont {P.~V.}\ \bibnamefont {Buividovich}},\ }\href {http://dx.doi.org/10.1016/j.nuclphysa.2014.02.022} {\bibfield  {journal} {\bibinfo  {journal} {Nucl.~Phys.~A}\ }\textbf {\bibinfo {volume} {925}},\ \bibinfo {pages} {218 } (\bibinfo {year} {2014})},\ \Eprint {https://arxiv.org/abs/1312.1843} {1312.1843} \BibitemShut {NoStop}%
\bibitem [{\citenamefont {Greif}\ \emph {et~al.}(2014)\citenamefont {Greif}, \citenamefont {Bouras}, \citenamefont {Xu},\ and\ \citenamefont {Greiner}}]{Greiner:1408.7049}%
  \BibitemOpen
  \bibfield  {author} {\bibinfo {author} {\bibfnamefont {M.}~\bibnamefont {Greif}}, \bibinfo {author} {\bibfnamefont {I.}~\bibnamefont {Bouras}}, \bibinfo {author} {\bibfnamefont {Z.}~\bibnamefont {Xu}},\ and\ \bibinfo {author} {\bibfnamefont {C.}~\bibnamefont {Greiner}},\ }\href {http://dx.doi.org/10.1103/PhysRevD.90.094014} {\bibfield  {journal} {\bibinfo  {journal} {Phys.~Rev.~D}\ }\textbf {\bibinfo {volume} {90}},\ \bibinfo {pages} {094014} (\bibinfo {year} {2014})},\ \Eprint {https://arxiv.org/abs/1408.7049} {1408.7049} \BibitemShut {NoStop}%
\bibitem [{\citenamefont {Puglisi}\ \emph {et~al.}(2014)\citenamefont {Puglisi}, \citenamefont {Plumari},\ and\ \citenamefont {Greco}}]{Puglisi:1408.7043}%
  \BibitemOpen
  \bibfield  {author} {\bibinfo {author} {\bibfnamefont {A.}~\bibnamefont {Puglisi}}, \bibinfo {author} {\bibfnamefont {S.}~\bibnamefont {Plumari}},\ and\ \bibinfo {author} {\bibfnamefont {V.}~\bibnamefont {Greco}},\ }\href {http://dx.doi.org/10.1103/PhysRevD.90.114009} {\bibfield  {journal} {\bibinfo  {journal} {Phys.~Rev.~D}\ }\textbf {\bibinfo {volume} {90}},\ \bibinfo {pages} {114009} (\bibinfo {year} {2014})},\ \Eprint {https://arxiv.org/abs/1408.7043} {1408.7043} \BibitemShut {NoStop}%
\bibitem [{\citenamefont {Qin}(2015)}]{Qin:1307.4587}%
  \BibitemOpen
  \bibfield  {author} {\bibinfo {author} {\bibfnamefont {S.}~\bibnamefont {Qin}},\ }\href {http://dx.doi.org/10.1016/j.physletb.2015.02.009} {\bibfield  {journal} {\bibinfo  {journal} {Phys.~Lett.~B}\ }\textbf {\bibinfo {volume} {742}},\ \bibinfo {pages} {358 } (\bibinfo {year} {2015})},\ \Eprint {https://arxiv.org/abs/1307.4587} {1307.4587} \BibitemShut {NoStop}%
\bibitem [{\citenamefont {Greif}\ \emph {et~al.}(2017)\citenamefont {Greif}, \citenamefont {Greiner},\ and\ \citenamefont {Denicol}}]{Greiner:1602.05085}%
  \BibitemOpen
  \bibfield  {author} {\bibinfo {author} {\bibfnamefont {M.}~\bibnamefont {Greif}}, \bibinfo {author} {\bibfnamefont {C.}~\bibnamefont {Greiner}},\ and\ \bibinfo {author} {\bibfnamefont {G.~S.}\ \bibnamefont {Denicol}},\ }\href {http://dx.doi.org/10.1103/PhysRevD.96.059902} {\bibfield  {journal} {\bibinfo  {journal} {Phys.~Rev.~D}\ }\textbf {\bibinfo {volume} {96}},\ \bibinfo {pages} {059902(E)} (\bibinfo {year} {2017})},\ \Eprint {https://arxiv.org/abs/1602.05085} {1602.05085} \BibitemShut {NoStop}%
\bibitem [{\citenamefont {{Fernandez-Fraile}}\ and\ \citenamefont {{Gomez Nicola}}(2006)}]{Fernandez:hep-ph/0512283}%
  \BibitemOpen
  \bibfield  {author} {\bibinfo {author} {\bibfnamefont {D.}~\bibnamefont {{Fernandez-Fraile}}}\ and\ \bibinfo {author} {\bibfnamefont {A.}~\bibnamefont {{Gomez Nicola}}},\ }\href {http://dx.doi.org/10.1103/PhysRevD.73.045025} {\bibfield  {journal} {\bibinfo  {journal} {Phys.~Rev.~D}\ }\textbf {\bibinfo {volume} {73}},\ \bibinfo {pages} {045025} (\bibinfo {year} {2006})},\ \Eprint {https://arxiv.org/abs/hep-ph/0512283} {hep-ph/0512283} \BibitemShut {NoStop}%
\bibitem [{\citenamefont {McLerran}\ and\ \citenamefont {Toimela}(1985)}]{McLerran:85:1}%
  \BibitemOpen
  \bibfield  {author} {\bibinfo {author} {\bibfnamefont {L.~D.}\ \bibnamefont {McLerran}}\ and\ \bibinfo {author} {\bibfnamefont {T.}~\bibnamefont {Toimela}},\ }\href {http://link.aps.org/doi/10.1103/PhysRevD.31.545} {\bibfield  {journal} {\bibinfo  {journal} {Phys.~Rev.~D}\ }\textbf {\bibinfo {volume} {31}},\ \bibinfo {pages} {545 } (\bibinfo {year} {1985})}\BibitemShut {NoStop}%
\bibitem [{\citenamefont {McLerran}\ and\ \citenamefont {Skokov}(929)}]{Skokov:1305.0774}%
  \BibitemOpen
  \bibfield  {author} {\bibinfo {author} {\bibfnamefont {L.}~\bibnamefont {McLerran}}\ and\ \bibinfo {author} {\bibfnamefont {V.}~\bibnamefont {Skokov}},\ }\href {http://dx.doi.org/10.1016/j.nuclphysa.2014.05.008} {\bibfield  {journal} {\bibinfo  {journal} {Nucl.~Phys.~A}\ }\textbf {\bibinfo {volume} {2014}},\ \bibinfo {pages} {184 – 190} (\bibinfo {year} {929})},\ \Eprint {https://arxiv.org/abs/1305.0774} {1305.0774} \BibitemShut {NoStop}%
\bibitem [{\citenamefont {Dubla}\ \emph {et~al.}(2020)\citenamefont {Dubla}, \citenamefont {G\"{u}rsoy},\ and\ \citenamefont {Snellings}}]{Gursoy:2009.09727}%
  \BibitemOpen
  \bibfield  {author} {\bibinfo {author} {\bibfnamefont {A.}~\bibnamefont {Dubla}}, \bibinfo {author} {\bibfnamefont {U.}~\bibnamefont {G\"{u}rsoy}},\ and\ \bibinfo {author} {\bibfnamefont {R.}~\bibnamefont {Snellings}},\ }\href@noop {} {\bibfield  {journal} {\bibinfo  {journal} {Mod.~Phys.~Lett.~A}\ }\textbf {\bibinfo {volume} {35}},\ \bibinfo {pages} {2050324} (\bibinfo {year} {2020})},\ \Eprint {https://arxiv.org/abs/2009.09727} {2009.09727} \BibitemShut {NoStop}%
\bibitem [{\citenamefont {Grieninger}\ \emph {et~al.}(2025)\citenamefont {Grieninger}, \citenamefont {Morales-Tejera},\ and\ \citenamefont {Romeu}}]{Grieninger:2503.10593}%
  \BibitemOpen
  \bibfield  {author} {\bibinfo {author} {\bibfnamefont {S.}~\bibnamefont {Grieninger}}, \bibinfo {author} {\bibfnamefont {S.}~\bibnamefont {Morales-Tejera}},\ and\ \bibinfo {author} {\bibfnamefont {P.~G.}\ \bibnamefont {Romeu}},\ }\href {https://doi.org/10.1103/xgkt-1qsb} {\bibfield  {journal} {\bibinfo  {journal} {Phys. Rev. D}\ }\textbf {\bibinfo {volume} {112}},\ \bibinfo {pages} {036003} (\bibinfo {year} {2025})},\ \Eprint {https://arxiv.org/abs/2503.10593} {arXiv:2503.10593 [hep-ph]} \BibitemShut {NoStop}%
\bibitem [{\citenamefont {Goswami}\ \emph {et~al.}(2015)\citenamefont {Goswami}, \citenamefont {Pixley},\ and\ \citenamefont {Das~Sarma}}]{Goswami:1503.02069}%
  \BibitemOpen
  \bibfield  {author} {\bibinfo {author} {\bibfnamefont {P.}~\bibnamefont {Goswami}}, \bibinfo {author} {\bibfnamefont {J.~H.}\ \bibnamefont {Pixley}},\ and\ \bibinfo {author} {\bibfnamefont {S.}~\bibnamefont {Das~Sarma}},\ }\href {https://doi.org/10.1103/PhysRevB.92.075205} {\bibfield  {journal} {\bibinfo  {journal} {Phys. Rev. B}\ }\textbf {\bibinfo {volume} {92}},\ \bibinfo {pages} {075205} (\bibinfo {year} {2015})},\ \Eprint {https://arxiv.org/abs/1503.02069} {arXiv:1503.02069 [cond-mat.mes-hall]} \BibitemShut {NoStop}%
\bibitem [{\citenamefont {Gorbar}\ \emph {et~al.}(2014)\citenamefont {Gorbar}, \citenamefont {Miransky},\ and\ \citenamefont {Shovkovy}}]{Gorbar:1312.0027}%
  \BibitemOpen
  \bibfield  {author} {\bibinfo {author} {\bibfnamefont {E.~V.}\ \bibnamefont {Gorbar}}, \bibinfo {author} {\bibfnamefont {V.~A.}\ \bibnamefont {Miransky}},\ and\ \bibinfo {author} {\bibfnamefont {I.~A.}\ \bibnamefont {Shovkovy}},\ }\href {https://doi.org/10.1103/PhysRevB.89.085126} {\bibfield  {journal} {\bibinfo  {journal} {Physical Review B}\ }\textbf {\bibinfo {volume} {89}},\ \bibinfo {pages} {085126} (\bibinfo {year} {2014})},\ \Eprint {https://arxiv.org/abs/1312.0027} {arXiv:1312.0027 [cond-mat.mes-hall]} \BibitemShut {NoStop}%
\bibitem [{\citenamefont {Meyer}(2011)}]{Meyer:1104.3708}%
  \BibitemOpen
  \bibfield  {author} {\bibinfo {author} {\bibfnamefont {H.~B.}\ \bibnamefont {Meyer}},\ }\href {http://dx.doi.org/10.1140/epja/i2011-11086-3} {\bibfield  {journal} {\bibinfo  {journal} {Eur.~Phys.~J.~A}\ }\textbf {\bibinfo {volume} {47}},\ \bibinfo {pages} {86} (\bibinfo {year} {2011})},\ \Eprint {https://arxiv.org/abs/1104.3708} {1104.3708} \BibitemShut {NoStop}%
\bibitem [{\citenamefont {Aarts}\ and\ \citenamefont {Nikolaev}(2021)}]{Nikolaev:2008.12326}%
  \BibitemOpen
  \bibfield  {author} {\bibinfo {author} {\bibfnamefont {G.}~\bibnamefont {Aarts}}\ and\ \bibinfo {author} {\bibfnamefont {A.}~\bibnamefont {Nikolaev}},\ }\href {https://dx.doi.org/10.1140/epja/s10050-021-00436-5} {\bibfield  {journal} {\bibinfo  {journal} {Eur.~Phys.~J.~A}\ }\textbf {\bibinfo {volume} {57}},\ \bibinfo {pages} {118} (\bibinfo {year} {2021})},\ \Eprint {https://arxiv.org/abs/2008.12326} {2008.12326} \BibitemShut {NoStop}%
\bibitem [{\citenamefont {Steinert}\ and\ \citenamefont {Cassing}(2014)}]{Cassing:1312.3189}%
  \BibitemOpen
  \bibfield  {author} {\bibinfo {author} {\bibfnamefont {T.}~\bibnamefont {Steinert}}\ and\ \bibinfo {author} {\bibfnamefont {W.}~\bibnamefont {Cassing}},\ }\href {http://dx.doi.org/10.1103/PhysRevC.89.035203} {\bibfield  {journal} {\bibinfo  {journal} {Phys.~Rev.~C}\ }\textbf {\bibinfo {volume} {89}},\ \bibinfo {pages} {035203} (\bibinfo {year} {2014})},\ \Eprint {https://arxiv.org/abs/1312.3189} {1312.3189} \BibitemShut {NoStop}%
\bibitem [{\citenamefont {Soloveva}\ \emph {et~al.}(2020)\citenamefont {Soloveva}, \citenamefont {Moreau},\ and\ \citenamefont {Bratkovskaya}}]{Bratkovskaya:1911.08547}%
  \BibitemOpen
  \bibfield  {author} {\bibinfo {author} {\bibfnamefont {O.}~\bibnamefont {Soloveva}}, \bibinfo {author} {\bibfnamefont {P.}~\bibnamefont {Moreau}},\ and\ \bibinfo {author} {\bibfnamefont {E.}~\bibnamefont {Bratkovskaya}},\ }\href {http://dx.doi.org/10.1103/PhysRevC.101.045203} {\bibfield  {journal} {\bibinfo  {journal} {Phys.~Rev.~C}\ }\textbf {\bibinfo {volume} {101}},\ \bibinfo {pages} {045203} (\bibinfo {year} {2020})},\ \Eprint {https://arxiv.org/abs/1911.08547} {1911.08547} \BibitemShut {NoStop}%
\bibitem [{\citenamefont {Tripolt}\ \emph {et~al.}(2019)\citenamefont {Tripolt}, \citenamefont {Jung}, \citenamefont {Tanji}, \citenamefont {{von Smekal}},\ and\ \citenamefont {Wambach}}]{Smekal:1807.04952}%
  \BibitemOpen
  \bibfield  {author} {\bibinfo {author} {\bibfnamefont {R.-A.}\ \bibnamefont {Tripolt}}, \bibinfo {author} {\bibfnamefont {C.}~\bibnamefont {Jung}}, \bibinfo {author} {\bibfnamefont {N.}~\bibnamefont {Tanji}}, \bibinfo {author} {\bibfnamefont {L.}~\bibnamefont {{von Smekal}}},\ and\ \bibinfo {author} {\bibfnamefont {J.}~\bibnamefont {Wambach}},\ }\href {http://dx.doi.org/10.1016/j.nuclphysa.2018.08.017} {\bibfield  {journal} {\bibinfo  {journal} {Nucl.~Phys.~A}\ }\textbf {\bibinfo {volume} {982}},\ \bibinfo {pages} {775 } (\bibinfo {year} {2019})},\ \Eprint {https://arxiv.org/abs/1807.04952} {1807.04952} \BibitemShut {NoStop}%
\bibitem [{\citenamefont {Kadam}\ \emph {et~al.}(2018)\citenamefont {Kadam}, \citenamefont {Mishra},\ and\ \citenamefont {Thakur}}]{Kadam:1712.03805}%
  \BibitemOpen
  \bibfield  {author} {\bibinfo {author} {\bibfnamefont {G.}~\bibnamefont {Kadam}}, \bibinfo {author} {\bibfnamefont {H.}~\bibnamefont {Mishra}},\ and\ \bibinfo {author} {\bibfnamefont {L.}~\bibnamefont {Thakur}},\ }\href {http://dx.doi.org/10.1103/PhysRevD.98.114001} {\bibfield  {journal} {\bibinfo  {journal} {Phys.~Rev.~D}\ }\textbf {\bibinfo {volume} {98}},\ \bibinfo {pages} {114001} (\bibinfo {year} {2018})},\ \Eprint {https://arxiv.org/abs/1712.03805} {1712.03805} \BibitemShut {NoStop}%
\bibitem [{\citenamefont {Ghosh}(2017)}]{Ghosh:1607.01340}%
  \BibitemOpen
  \bibfield  {author} {\bibinfo {author} {\bibfnamefont {S.}~\bibnamefont {Ghosh}},\ }\href {http://dx.doi.org/10.1103/PhysRevD.95.036018} {\bibfield  {journal} {\bibinfo  {journal} {Phys.~Rev.~D}\ }\textbf {\bibinfo {volume} {95}},\ \bibinfo {pages} {036018} (\bibinfo {year} {2017})},\ \Eprint {https://arxiv.org/abs/1607.01340} {1607.01340} \BibitemShut {NoStop}%
\bibitem [{\citenamefont {Ulybyshev}\ \emph {et~al.}(2017)\citenamefont {Ulybyshev}, \citenamefont {Winterowd},\ and\ \citenamefont {Zafeiropoulos}}]{Ulybyshev:1707.04212}%
  \BibitemOpen
  \bibfield  {author} {\bibinfo {author} {\bibfnamefont {M.}~\bibnamefont {Ulybyshev}}, \bibinfo {author} {\bibfnamefont {C.}~\bibnamefont {Winterowd}},\ and\ \bibinfo {author} {\bibfnamefont {S.}~\bibnamefont {Zafeiropoulos}},\ }\href {http://dx.doi.org/10.1103/PhysRevB.96.205115} {\bibfield  {journal} {\bibinfo  {journal} {Phys.~Rev.~B}\ }\textbf {\bibinfo {volume} {96}},\ \bibinfo {pages} {205115} (\bibinfo {year} {2017})},\ \Eprint {https://arxiv.org/abs/1707.04212} {1707.04212} \BibitemShut {NoStop}%
\bibitem [{\citenamefont {Ding}\ \emph {et~al.}(2011)\citenamefont {Ding}, \citenamefont {Francis}, \citenamefont {Kaczmarek}, \citenamefont {Karsch}, \citenamefont {Laermann},\ and\ \citenamefont {Soeldner}}]{Kaczmarek:1012.4963}%
  \BibitemOpen
  \bibfield  {author} {\bibinfo {author} {\bibfnamefont {H.}~\bibnamefont {Ding}}, \bibinfo {author} {\bibfnamefont {A.}~\bibnamefont {Francis}}, \bibinfo {author} {\bibfnamefont {O.}~\bibnamefont {Kaczmarek}}, \bibinfo {author} {\bibfnamefont {F.}~\bibnamefont {Karsch}}, \bibinfo {author} {\bibfnamefont {E.}~\bibnamefont {Laermann}},\ and\ \bibinfo {author} {\bibfnamefont {W.}~\bibnamefont {Soeldner}},\ }\href {http://dx.doi.org/10.1103/PhysRevD.83.034504} {\bibfield  {journal} {\bibinfo  {journal} {Phys.~Rev.~D}\ }\textbf {\bibinfo {volume} {83}},\ \bibinfo {pages} {034504} (\bibinfo {year} {2011})},\ \Eprint {https://arxiv.org/abs/1012.4963} {1012.4963} \BibitemShut {NoStop}%
\bibitem [{\citenamefont {Fukushima}\ and\ \citenamefont {Hidaka}(2020)}]{Fukushima:1906.02683}%
  \BibitemOpen
  \bibfield  {author} {\bibinfo {author} {\bibfnamefont {K.}~\bibnamefont {Fukushima}}\ and\ \bibinfo {author} {\bibfnamefont {Y.}~\bibnamefont {Hidaka}},\ }\href {https://doi.org/10.1007/JHEP04(2020)162} {\bibfield  {journal} {\bibinfo  {journal} {JHEP}\ }\textbf {\bibinfo {volume} {04}},\ \bibinfo {pages} {162}},\ \Eprint {https://arxiv.org/abs/1906.02683} {arXiv:1906.02683 [hep-ph]} \BibitemShut {NoStop}%
\bibitem [{\citenamefont {Shaikh}\ \emph {et~al.}(2023)\citenamefont {Shaikh}, \citenamefont {Rath}, \citenamefont {Dash},\ and\ \citenamefont {Panda}}]{Shaikh:2210.15388}%
  \BibitemOpen
  \bibfield  {author} {\bibinfo {author} {\bibfnamefont {A.}~\bibnamefont {Shaikh}}, \bibinfo {author} {\bibfnamefont {S.}~\bibnamefont {Rath}}, \bibinfo {author} {\bibfnamefont {S.}~\bibnamefont {Dash}},\ and\ \bibinfo {author} {\bibfnamefont {B.}~\bibnamefont {Panda}},\ }\href {https://doi.org/10.1103/PhysRevD.108.056021} {\bibfield  {journal} {\bibinfo  {journal} {Phys. Rev. D}\ }\textbf {\bibinfo {volume} {108}},\ \bibinfo {pages} {056021} (\bibinfo {year} {2023})},\ \Eprint {https://arxiv.org/abs/2210.15388} {arXiv:2210.15388 [hep-ph]} \BibitemShut {NoStop}%
\bibitem [{\citenamefont {Thakur}\ and\ \citenamefont {Srivastava}(2019)}]{Thakur:1910.12087}%
  \BibitemOpen
  \bibfield  {author} {\bibinfo {author} {\bibfnamefont {L.}~\bibnamefont {Thakur}}\ and\ \bibinfo {author} {\bibfnamefont {P.~K.}\ \bibnamefont {Srivastava}},\ }\href {https://doi.org/10.1103/PhysRevD.100.076016} {\bibfield  {journal} {\bibinfo  {journal} {Phys. Rev. D}\ }\textbf {\bibinfo {volume} {100}},\ \bibinfo {pages} {076016} (\bibinfo {year} {2019})},\ \Eprint {https://arxiv.org/abs/1910.12087} {arXiv:1910.12087 [hep-ph]} \BibitemShut {NoStop}%
\bibitem [{\citenamefont {Skokov}\ \emph {et~al.}(2009)\citenamefont {Skokov}, \citenamefont {Illarionov},\ and\ \citenamefont {Toneev}}]{Skokov:0907.1396}%
  \BibitemOpen
  \bibfield  {author} {\bibinfo {author} {\bibfnamefont {V.}~\bibnamefont {Skokov}}, \bibinfo {author} {\bibfnamefont {A.}~\bibnamefont {Illarionov}},\ and\ \bibinfo {author} {\bibfnamefont {V.}~\bibnamefont {Toneev}},\ }\href {https://doi.org/10.1142/S0217751X09047570} {\bibfield  {journal} {\bibinfo  {journal} {Int. J. Mod. Phys. A}\ }\textbf {\bibinfo {volume} {24}},\ \bibinfo {pages} {5925} (\bibinfo {year} {2009})},\ \Eprint {https://arxiv.org/abs/0907.1396} {0907.1396} \BibitemShut {NoStop}%
\bibitem [{\citenamefont {Tuchin}(2013)}]{Tuchin:1305.5806}%
  \BibitemOpen
  \bibfield  {author} {\bibinfo {author} {\bibfnamefont {K.}~\bibnamefont {Tuchin}},\ }\href {https://doi.org/10.1103/PhysRevC.88.024911} {\bibfield  {journal} {\bibinfo  {journal} {Phys. Rev. C}\ }\textbf {\bibinfo {volume} {88}},\ \bibinfo {pages} {024911} (\bibinfo {year} {2013})},\ \Eprint {https://arxiv.org/abs/1305.5806} {arXiv:1305.5806 [hep-ph]} \BibitemShut {NoStop}%
\end{thebibliography}


\end{document}